\documentclass[twocolumn,trackchanges]{aastex631}

\newcommand{\e}{\mbox{e}^}
\newcommand{\amm}{NH$_3$}

\newcommand{\namm}{$N$(NH$_3$)}

\newcommand{\nh}{N(H$_2$)}
\newcommand{\kms}{km~s$^{-1}$}

\newcommand{\vlsr}{$v_{\mathrm{LSR}}$}
\newcommand{\sigv}{$\sigma_v$}
\newcommand{\signt}{$\sigma_\mathrm{NT}$}
\newcommand{\tkin}{$T_{K}$}

\newcommand{\msun}{M$_\odot$}

\newcommand{\xamm}{$X$(NH$_3$)}
\newcommand{\kmspc}{km~s$^{-1}$~pc$^{-1}$}
\newcommand{\sigps}{$\Sigma_\mathrm{PS}$}
\renewcommand{\edit}[1]{}

\accepted{April 10, 2024}

\submitjournal{ApJ}

\shorttitle{Stability of dense cores near the Serpens South protocluster}
\shortauthors{Friesen \& Jarvis}

\begin{document}

\title{The stability of dense cores near the Serpens South protocluster}

\correspondingauthor{Rachel Friesen}
\email{rachel.friesen@utoronto.ca}

\author[0000-0001-7594-8128]{Rachel K. Friesen}
\affiliation{David A. Dunlap Department of Astronomy \& Astrophysics \\
University of Toronto \\
50 St. George St., Toronto ON Canada}

\author[0009-0006-5612-7336]{Emma Jarvis}
\affiliation{David A. Dunlap Department of Astronomy \& Astrophysics \\
University of Toronto \\
50 St. George St., Toronto ON Canada}



\begin{abstract}

Most stars form in clusters and groups rather than in isolation.
We present $\lesssim 5$\arcsec\ angular resolution ($\sim 2000$~au, or 0.01~pc) Very Large Array \amm\ (1,1), (2,2), and (3,3) and 1.3~cm continuum emission observations of the dense gas within the Serpens South protocluster and extended filaments to the north and south.
\edit1{We identify 94 dense cores using a dendrogram analysis of the \amm\ (1,1) integrated intensity.}
Gas temperatures \tkin\ and non-thermal linewidths \signt\ both increase towards the centre of the young stellar cluster, in the dense gas generally and in the cores specifically.
We find that most cores \edit1{(54\%)} are super-virial, with gravitationally bound cores located primarily in the filaments.
\edit1{Cores in the protocluster have higher virial parameters by a factor $\sim 1.7$, driven primarily by the increased core \signt\ values. }
These cores cannot collapse to form stars unless they accrete additional mass or their core internal motions are reduced.
\edit1{The southern filament shows a significant velocity gradient previously interpreted as mass flow toward the cluster. 
We find more complex kinematics in the northern filament. }
We find a strong correlation between \signt\ and \tkin, and argue that the enhanced temperatures and non-thermal motions are due to mechanical heating and interaction between the protocluster-driven outflows and the dense gas. 
Filament-led accretion may also \edit1{contribute to} the increased \signt\ values.
Assuming a constant fraction of core mass ends up in the young stars, future star formation in the Serpens South protocluster will shift to higher masses by a factor $\sim 2$.

\end{abstract}



\section{Introduction} \label{sec:intro}

Star formation occurs due to the collapse of dense molecular clouds in the interstellar medium. Most stars form in groups or clusters within these clouds rather than in isolation \citep{lada_2003}. Many clustered star-forming regions share similar morphologies where filaments extend from a central hub that contains substantial mass and active star formation \citep{myers_2009}. These filaments can feed in additional mass to the central cluster as it is forming \citep{peretto_2013,kirk_2013}. Gravitationally bound cores that are likely to form additional stars are mostly found along these filaments \citep{konyves_2015}. 

Young stellar clusters have a large number of protostars in their centres. Protostellar heating has many impacts on the surrounding dense gas. Outflows from young stars can impact star formation because they can interact with surrounding gas \citep{fuller_2002}, limit star formation efficiency \citep{matzner_2000}, facilitate the accretion of material onto protostars, heat the gas, and inject energy and momentum into the surrounding cloud \citep{bachiller_1996}. The same cloud beside a stellar cluster could get warmed up, changing the balance of forces which can increase the masses of the next generation of stars \citep{hatchell_2013}. Outflows from low-mass protostars can affect their surrounding environment as they typically extend to distances of 0.1--1pc \citep{arce_2007}. They may feed turbulent motions that affect ongoing star formation \citep{Nakamura_2014}. Class 0 objects (the youngest, embedded protostars) in particular are the origin of powerful ejections of matter as they are in their main accretion phase \citep{arce_2007}. 

In order to study the role of a young protostellar cluster on surrounding dense gas and future generations of stars we present observations of the Serpens South cluster and surrounding filaments. Discovered in 2008 \citep{gutermuth_2008}, Serpens South is a nearby star forming region comprised of low and intermediate mass young stellar objects and is one of only a few active star forming regions located at a distance less than 500~pc. 
Serpens South consists of an infrared-bright protostellar cluster embedded within a hub-filament structure contained within a large complex of infrared-dark gas. \citet{gutermuth_2008} find that the cluster has a large fraction (77\%) of Class I protostars; young stars in the earliest stage of stellar evolution with a core not yet hot enough for nuclear fusion. Since its discovery many more young stellar objects have been identified \citep[e.g.,][]{dunham_2015, plunkett_2018, sun_2022}. This large fraction of protostars indicates that star formation in this region has only begun recently (($2 \pm 1)\times 10^5$ yr) and that the cluster has a young age. The Serpens South cluster also has a high star formation rate (90~M$_\odot$~Myr$^{-1}$) and contains sufficient mass to form new stars in the future. The main filaments contain $\sim 1660$~\msun\ \citep{friesen_2016}. This combination suggests that many more stars may yet form in this region, making it an ideal target to study the impact of clustered, low- to intermediate-mass star formation on the next generation of stars. 

Several studies have presented observations of filamentary accretion in Serpens South and outflows from it's protocluster. Narrow filaments are seen in the dense gas surrounding the protostellar cluster \citep{fernandez_2014}. Filamentary accretion flows may be an important mechanism for supplying material to the Serpens South protocluster \citep{kirk_2013}. \citet{Nakamura_2014} argue that the Serpens South central protocluster formation was triggered by filament collisions. \citet{plunkett_2015_outflows} find that outflows from the Serpens South protocluster contribute enough momentum and energy to maintain turbulence and that the estimated mass accretion rate is comparable to the lower limit infall rate onto the filament. \citet{Nakamura_2011} also suggest that the energy injected into the central embedded cluster by outflows may be enough to maintain the current level of non-thermal motion.

There has been controversy over the distance to Serpens South, with early values ranging from 260~pc \citep{gutermuth_2008} to $\sim 430$~pc \citep[e.g.,][]{dzib_2010,ortiz_leon_2018,zucker_2019}. \citet{ortiz_leon_2023} show conclusively via astrometry of H$_2$O masers toward an embedded cluster protostar, CARMA-6, that the cluster lies at a distance $d = 440.7 \pm 4.6$~pc, and we use this measurement in this paper. 

In this paper we investigate the structure and stability of the dense molecular gas of the Serpens South protocluster. We present high spatial resolution observations of the \amm\ (1,1) and (2,2) inversion transitions made with the Karl G. Jansky Very Large Array (VLA). \amm\ can be detected easily in quiescent dark clouds and regions of active star formation. \amm\ (1,1) and \amm\ (2,2) have low energy levels and are best tracers of cold (10 -- 25 K) gas with number densities $n > 10^3~\mathrm{cm}^{-3}$ \citep{shirley_2015}. Unlike continuum emission from dust, observations of \amm\ can probe the kinematics of dense gas. By measuring different inversion transitions we can obtain a direct measurement of the kinetic gas temperature \citep{ho_townes_1983} and gas velocity dispersion, which is required to determine if structures are gravitationally bound. Furthermore, \amm\ suffers less from freeze-out onto dust grains than carbon-based molecules such as CO \citep{difrancesco_2007}. For these reasons, \amm\ is an ideal tracer to study the structure and kinematics of the dense gas in Serpens South.

In Section \ref{section:observations} we describe our observations, data reduction and line fitting. In Section \ref{section:results} we present the resulting integrated intensity and parameter maps obtained from the hyperfine fitting. In Section \ref{section:Analysis} we describe the results of the dendrogram analysis of the \amm\ (1,1) integrated intensity map used to identify the locations of dense cores and describe the method used to calculate the core masses. We also determine the stability of the cores. We then describe the large-scale properties of the dense gas. In Section \ref{section:discussion} we discuss the stability of cores, the influence of the cluster on the surrounding dense gas and the future of star formation in Serpens South. In Section \ref{section:summary} we provide a summary of our results.

\begin{figure*}
    \includegraphics[width=0.48\textwidth]{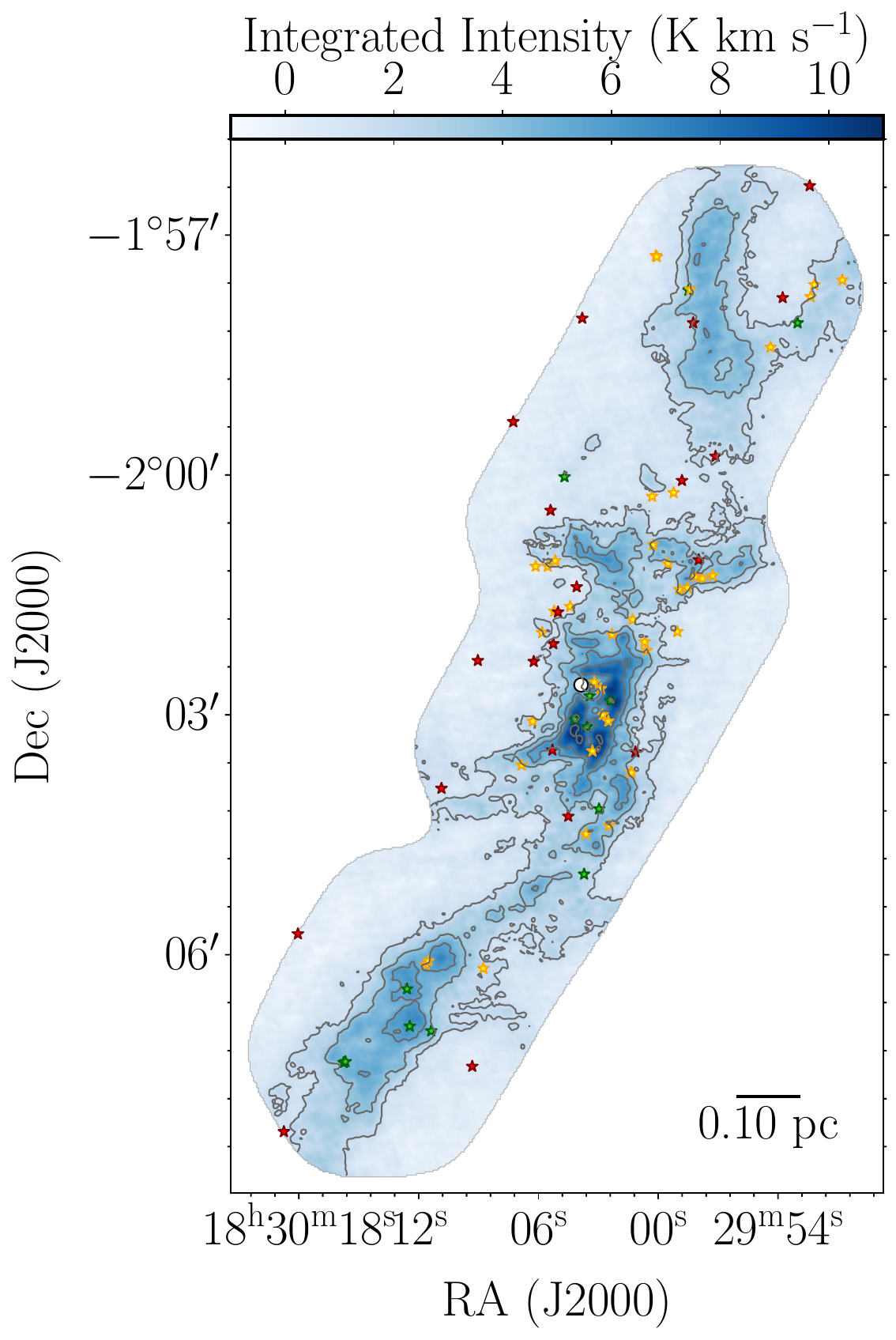}
    \label{fig:NH3_11_Protostars}
    \includegraphics[width=0.48\textwidth]{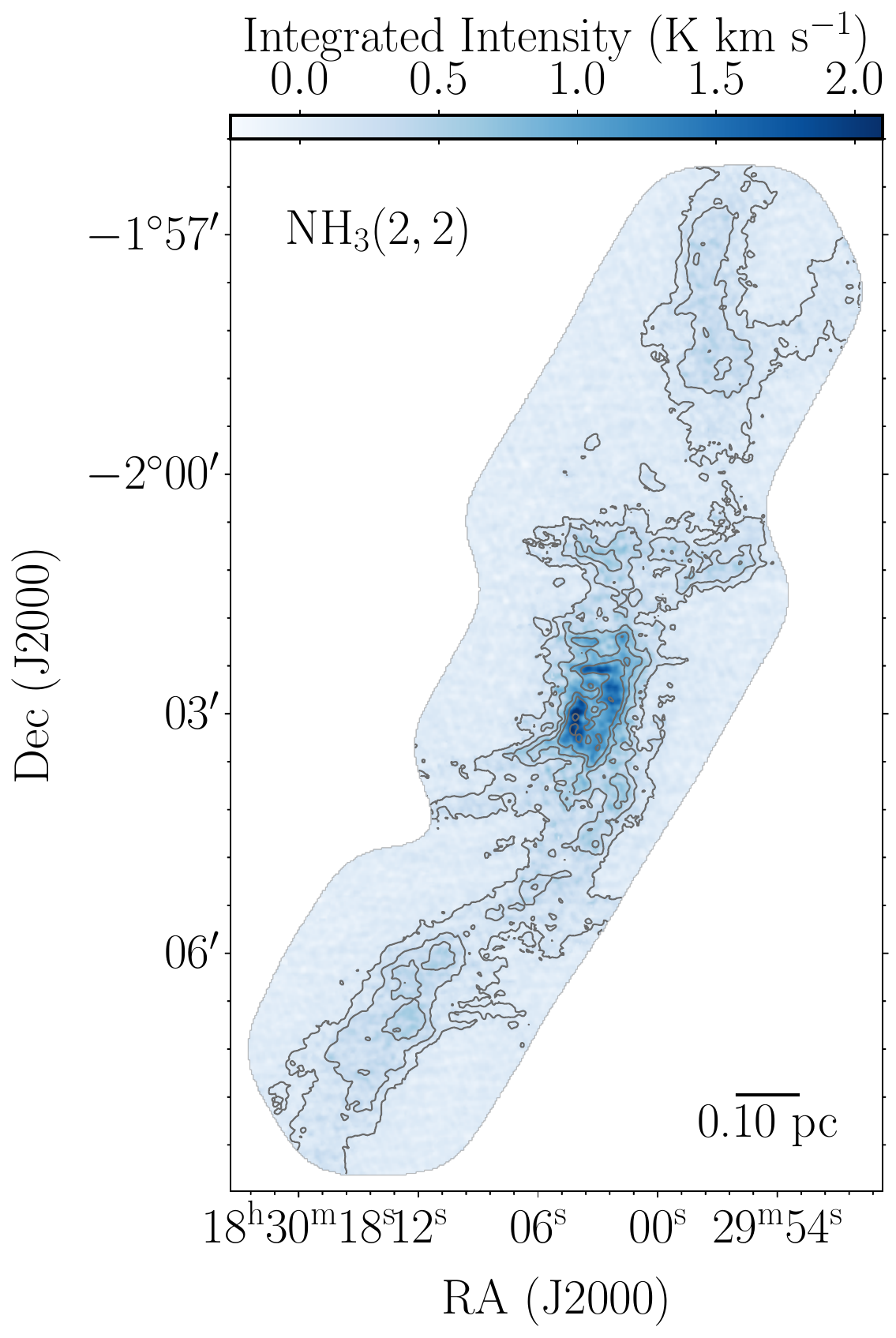}
    \caption{Left: Map of the \amm\ (1,1) integrated intensity towards Serpens South with the locations of YSOs. Green, yellow and red stars mark the locations of Class 0, Class I and Class II protostars, respectively. Flat spectrum sources are also marked by yellow stars. These YSO locations are obtained by merging the catalogs from \citet{sun_2022} and \citet{pokhrel_2023}. The white circle marks the location of the centre of the protostar distribution in the cluster obtained from \citet{sun_2022b}. Right: Map of the \amm\ (2,2) integrated intensity towards Serpens South. Overlayed contours on both plots are of the \amm\ (1,1) integrated intensity data and begin at 1.8 K \kms\ and increase in 1.8 K \kms\ intervals. }\label{fig:protostars}
    
\end{figure*}

\section{Very Large Array mosaic observations and data reduction}\label{section:observations}

\subsection{\amm\ line emission}
\label{section:line_emission}

The Serpens South protocluster and extended filamentary structure to the northwest and southeast was observed as a mosaic of 36 individual pointed observations with the Very Large Array (VLA) in \amm(1,1), (2,2) and (3,3) emission from July 2015 to September 2015 \edit1{(VLA project ID 14A-180)}. The mosaic coverage was based on single dish \amm\ data of the entire Serpens South cloud \citep{friesen_2016}, and extends across the central stellar cluster and the north and south-facing filamentary structures identified in \amm.
The pointings were Nyquist-spaced by 61\arcsec\ ($1/\sqrt{3}$ times the primary beam at the rest frequency of \amm(1,1), $\nu_0 = 23.692$~GHz). The mosaic was divided into three separate groups of pointings, and scheduling blocks cycled through the groups on different dates and times, allowing for improved $uv$-coverage of the full mosaic while minimizing overhead. As a result, however, the overall rms noise is not completely uniform within the central mosaic where pointings overlap, as described further below. 

Bandpass and flux calibration were performed on 3C286, and J1851+0035 was used as the phase calibrator. Some data were flagged based on excessive noise or other considerations. Calibration of the line data was performed in CASA 5.4 following standard processes, and the continuum was subtracted from the line spectral windows. 

The \amm(1,1) and (2,2) data were then imaged in CASA using \texttt{tclean}. For each transition, we used single-dish data from the Green Bank Telescope \citep{friesen_2016} as a starting clean model. To do this, an initial mosaic was made of the VLA data alone using \texttt{tclean}, without any clean iterations, but with the desired final spatial resolution and a spectral resolution of 0.15~km~s$^{-1}$ to match the limits of the GBT data. A small taper was applied in the VLA imaging to produce a final angular resolution of $\sim 5$\arcsec\ to improve the signal-to-noise ratio (SNR). The GBT data were then rescaled from $T_{mb}$ to Jy~beam$^{-1}$ units and regridded to match the initial VLA mosaic along all three axes (position-position-velocity, or PPV). A clean mask at each spectral channel was created based on the observed emission in the GBT data. Deconvolution and imaging was then performed using \texttt{tclean} with the GBT data as the initial clean model, with the clean mask applied as described. An initial interactive clean was initially run and the clean mask revised as needed, after which clean was run in non-interactive mode, using multiscale deconvolution, to a threshold of 20~mJy~beam$^{-1}$. Finally, the single dish and cleaned interferometer cubes were combined via the \texttt{feather} task. The combined data were convolved to the angular resolution of the single-dish data as a check to ensure the combined fluxes were reasonable. The final combined maps were rescaled to units of $T_{mb}$ (K) for the line analysis described below. 

The final angular resolution of the combined \amm\ (1,1) and (2,2) data is 4\farcs8 $\times$ 4\farcs6, with a position angle of 80$^\circ$, corresponding to a spatial resolution of $\sim 2100$~au (0.01~pc) at the assumed distance to Serpens South. The sensitivity varies slightly over the map, from 0.13~K to 0.15~K in a 0.15~\kms\ spectral channel in the southeast, with the larger values toward the northern end of the mosaic. Since the maps combined both interferometric and single-dish data, they are sensitive to emission on all spatial scales. 

For \amm\ (3,3), the same imaging procedure was followed, but the VLA data were imaged with a spectral resolution of 0.4~\kms\ to better match the broad width of the detected lines, and to improve the SNR. The GBT data were also Hanning smoothed to a velocity resolution of 0.4~\kms\ before the regridding step and combination with the VLA data. The resulting angular resolution for the \amm\ (3,3) data are 4\farcs8 $\times$ 4\farcs6, with a position angle of 80$^\circ$. The sensitivity per 0.4~\kms\ velocity channel is $\sim 0.08$~K in $T_{mb}$ units in the mosaic overlap region. 

\begin{figure*}
    \centering
    \includegraphics[trim={0 5 10 0}, clip, width=0.373\textwidth]{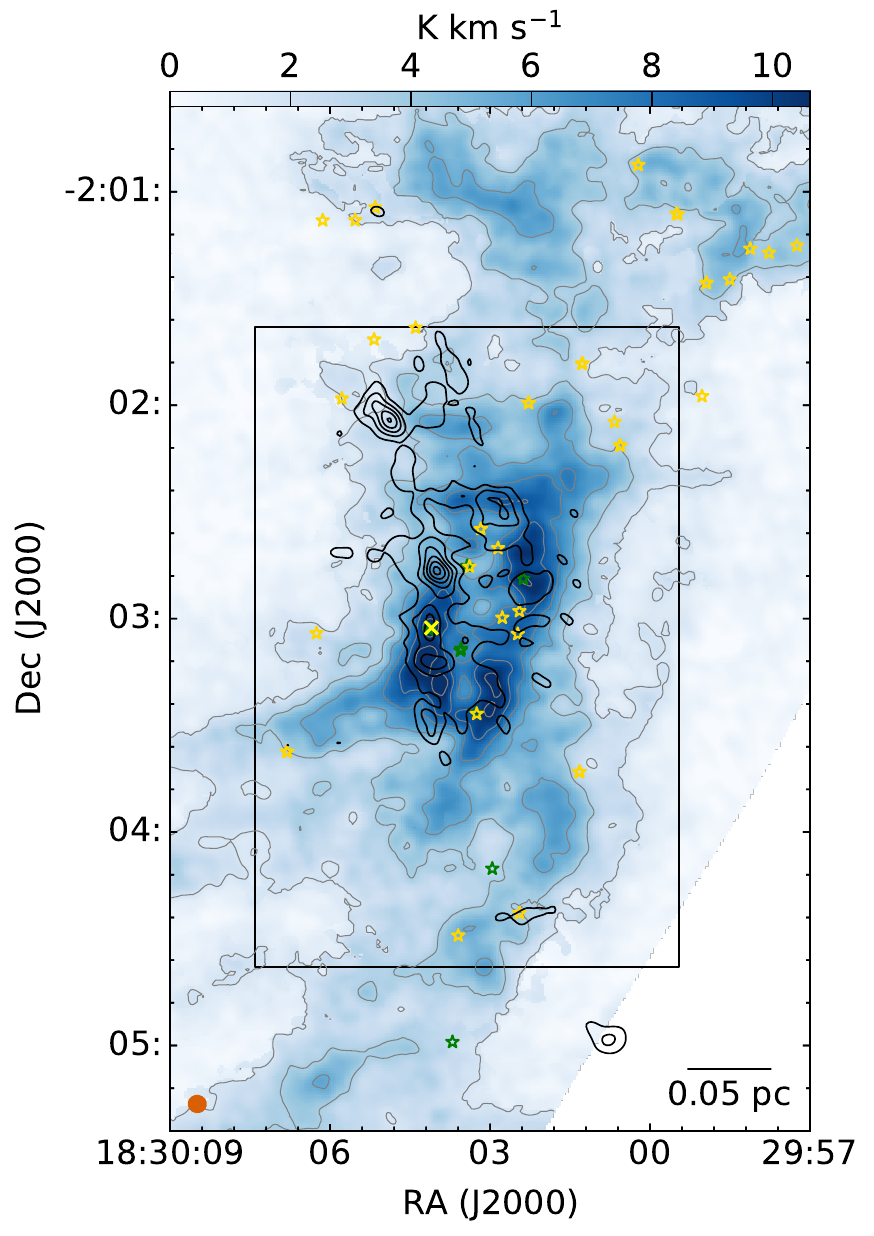}
    \hspace{0.1cm}
    \includegraphics[trim={70 0 120 0}, clip, width=0.35\textwidth]{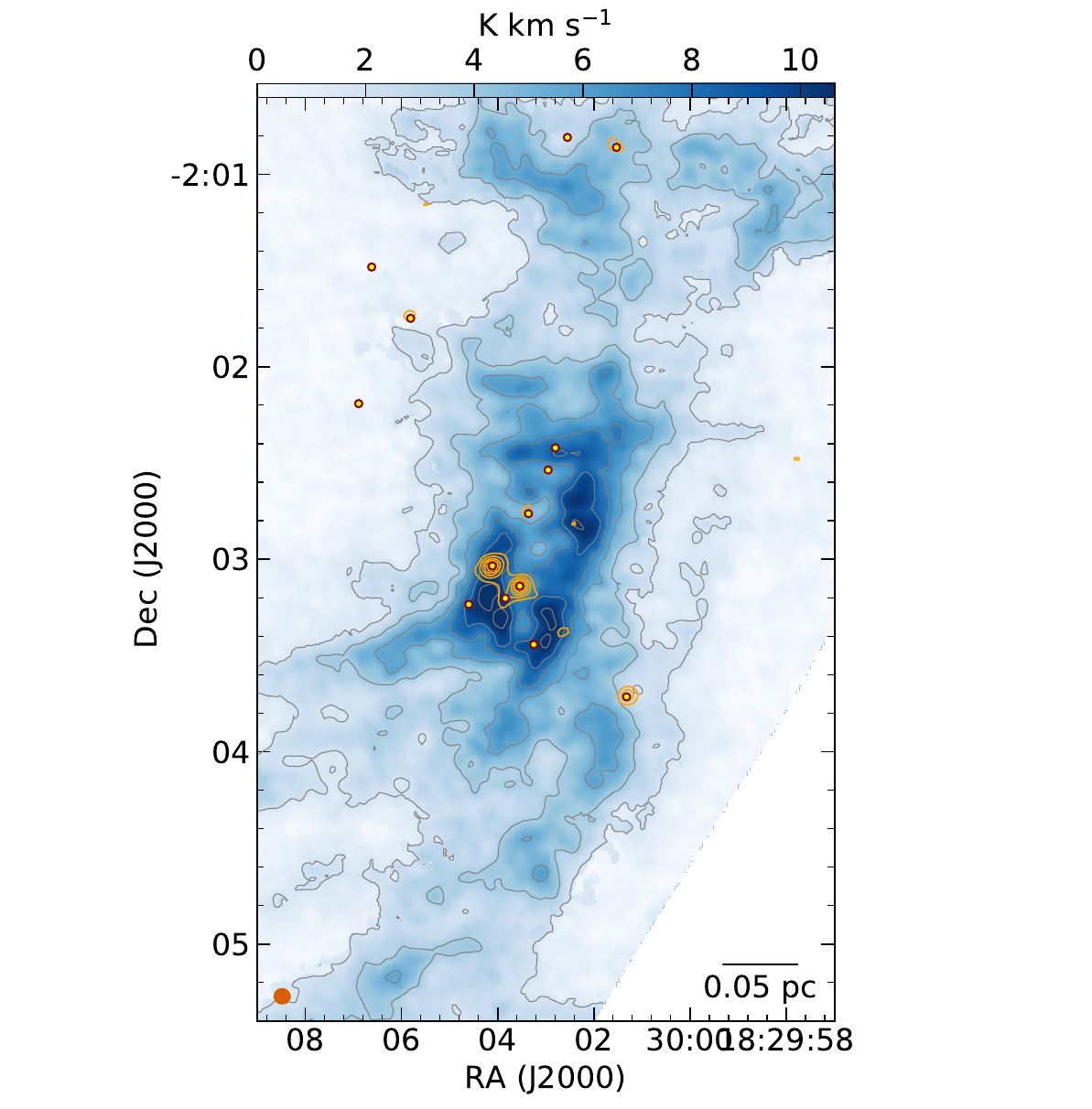}
    \caption{Left: Colorscale shows \amm\ (1,1) integrated intensity as in Figure \ref{fig:protostars}, zoomed in on the central YSO cluster. Grey contours show \amm(1,1) integrated intensity as in Figure \ref{fig:protostars}. Black contours show \amm\ (3,3) integrated intensity, starting at 4-$\sigma$ and increasing in 3-$\sigma$ increments. 
    Class 0, I, and flat spectrum sources are shown following Figure \ref{fig:NH3_11_Protostars}\edit1{, with the yellow `x' highlighting the Class 0/I source CARMA-7 \citep{plunkett_2015_outflows}}. The black box highlights the region used to compare gas properties near the cluster centre in Section \ref{sec:kinematics}. Right: Colorscale shows \amm\ (1,1) integrated intensity over the same region as at left. Orange contours identify 1.3~cm continuum emission beginning at 4-$\sigma$ and increasing in 3-$\sigma$ increments. Circles identify radio continuum sources from \citet{kern_2016}. The two brightest centimeter continuum sources correspond to previously identified sources at 4.75~GHz and 7.25~GHz \citep[VLA 12 and VLA 13;][]{kern_2016}.} 
    \label{fig:cont_33}
\end{figure*}

\subsubsection{\amm\ line fitting and moment maps}
\label{sec:line_fitting}

Across the mosaic, \amm(1,1) and (2,2) lines were fit jointly at each pixel following the method described in \citet{GAS_DR1}, which uses the cold\_ammonia spectral line fitter in \texttt{pyspeckit} \citep{pyspeckit_2011}. The model assumes that both the \amm(1,1) and (2,2) lines, and all the associated hyperfine components, are described by the same line-of-sight velocity in the Local Standard of Rest (LSR) frame, \vlsr, velocity dispersion \sigv, and excitation temperature $T_{ex}$. At each pixel with sufficient SNR, the fits return the \vlsr, \sigv, $T_{ex}$, as well as the gas kinetic temperature $T_K$, and the total \amm\ column density, $N($\amm$)$, along with uncertainties in each parameter based on the fit and the rms noise in the data. We further mask the resulting parameter maps based on the parameter uncertainties at each pixel. 

The integrated intensity maps for \amm\ (1,1) and (2,2) were then made following \citet{GAS_DR1}. Here, the line emission is integrated over the \amm\ hyperfine components, and the included spectral windows are defined from the \vlsr\ and \sigv\ fit results. For pixels without good \amm\ fits, the mean \vlsr\ and \sigv\ were used from nearby fit results to determine the spectral regions over which to integrate. Where no fit results are available, the integrated intensity was calculated over spectral windows corresponding to the mean \vlsr\ and \sigv\ of the entire map. The rms noise was determined in the corresponding non-line spectral windows. We show the resulting \amm\ (1,1) and (2,2) integrated intensity maps in Figure \ref{fig:protostars}. 

The \amm\ (3,3) emission is an ortho-\amm\ transition, rather than para-\amm\ as for the (1,1) and (2,2) lines. A visual examination showed that the \amm\ (3,3) emission was clearly offset in position relative to the (1,1) and (2,2) lines, and shows different kinematics. As a result, we fit these data separately with a single Gaussian component, returning the line amplitude, \vlsr, and \sigv.

\subsection{Centimeter continuum emission}
\label{sec:continuum}

Simultaneously with the line observations, the WIDAR correlator was configured to have eight spectral windows (spw) of 128~MHz in width and 128 spectral channels each between 22.0~GHz and 23.4~GHz for continuum observations. The continuum spw central frequencies were chosen to avoid potential strong line emission. The data were calibrated with the VLA calibration pipeline (version 4.7.0). The spectral windows were checked for line emission, particularly at the pointing corresponding to the central stellar cluster where line emission is strongest, and several channels were flagged. The data were then imaged in \texttt{tclean}, using multi-frequency synthesis to produce a single continuum emission map. The same taper was used as for the line data, producing a final synthesized beam of 4\farcs9 $\times$ 4\farcs4 at a position angle of -39$^\circ$. The image rms noise is 34~$\mu$Jy~beam$^{-1}$. 

Toward Serpens South, centimeter radio continuum emission is primarily detected as compact sources within the protocluster core. We show in Figure \ref{fig:cont_33} (right) a zoomed-in view of the \amm\ (1,1) integrated intensity toward the cluster center, overlaid with contours highlighting 1.3~cm continuum sources with SNR $> 4$. We identify a 1.3~cm source by requiring at least a 4-$\sigma$ detection in the continuum data with a corresponding source previously identified at radio or submillimeter wavelengths \citep[e.g.,][]{kern_2016,plunkett_2018}. Markers show 7.25~GHz (4.14~cm) and 4.75~GHz (6.31~cm) radio continuum sources \citep{kern_2016}. The strongest continuum detections correspond to sources VLA 12 and VLA 13 in \citeauthor{kern_2016}, which are also detected at (sub)millimeter wavelengths \citep[CARMA-7 and CARMA-6, respectively;][]{plunkett_2015_outflows,plunkett_2018}, and are identified as Class 0/I protostars \citep{dunham_2015}. 
In our data, these sources are extended. 
At this wavelength, the continuum emission is likely a mix of both thermal dust emission and thermal bremsstrahlung free-free emission. A list of continuum sources and their properties is given in Table \ref{tab:continuum}, but we do not analyse the continuum data further in this paper. 

\section{Results}\label{section:results}

\subsection{Distribution of NH$_\textbf{3}$ emission at 2000~au in Serpens South}

\subsubsection{\amm\ moment maps}\label{section:moment_maps}

Figure \ref{fig:protostars} shows the \amm\ (1,1) and (2,2) integrated intensity maps towards Serpens South. Both images are overlaid with \amm\ (1,1) integrated intensity contours in intervals of 1.8~K~\kms. \amm\ (1,1) is detected over most of the map, while \amm\ (2,2) is well-detected over a smaller area than \amm\ (1,1). For both \amm\ (1,1) and \amm\ (2,2), we find strong line emission towards the central cluster and and in the connected filaments towards the northwest and southeast of the central cluster. There is also faint \amm\ (1,1) emission in the regions between the central cluster and filaments. 

Figure \ref{fig:protostars} also shows the locations of 77 known or candidate YSOs in the Serpens South region. Of these, 15 are classified as Class 0, 36 are classified as Class I, 5 are classified as flat-spectrum and 21 are classified as Class II protostars. We obtain this catalog of protostars from \citet{sun_2022} and \citet{pokhrel_2023}. \citet{sun_2022} identify YSOs using deep near-infrared observations of CFHT in combination with 2MASS, UKIDSS, and Spitzer catalogs.  \citet{pokhrel_2023} classify YSOs by their SEDs using 1--850~$\mu$m observations assembled from  2MASS, Spitzer, Herschel, the Wide-field Infrared Survey Explorer, and James Clerk Maxwell Telescope/SCUBA-2 data. For sources that appear in both catalogs, we use the classification from \citet{pokhrel_2023} as they include data at longer wavelengths. 

Figure \ref{fig:cont_33} (left) shows a zoomed-in view of the \amm\ (1,1) integrated intensity toward the central cluster, overlaid with contours highlighting the \amm\ (3,3) integrated intensity. \amm\ (3,3) is only detected within the area shown in the Figure, and its distribution is offset from the \amm\ (1,1) and (2,2) emission. Strong \amm\ (3,3) emission is \edit1{coincident with} the collimated molecular outflow originating near the millimeter source CARMA-7 \citep{plunkett_2015_epi,plunkett_2018}, also detected as a radio continuum source \edit1{here and in previous observations} \citep[VLA 12;][]{kern_2016}.
\edit1{We discuss in more detail the correspondence between the \amm\ (3,3) emission and outflows in Section \ref{section:cluster_effect}.}

\begin{figure}
    \centering
    \includegraphics[width=0.45\textwidth]{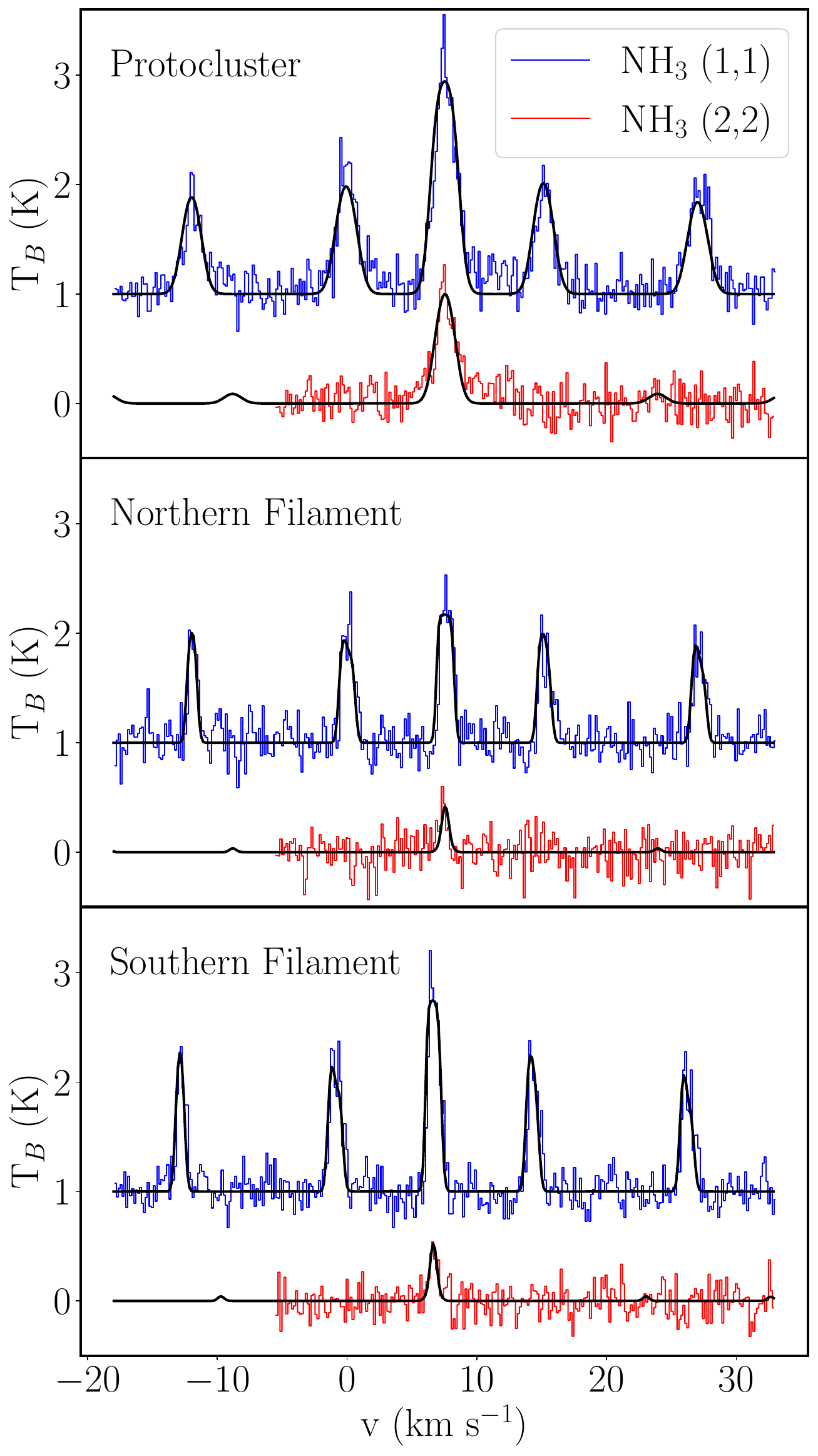}
    \caption{Spectra towards the emission peak in the Serpens South protocluster (top), in the filament extending north-west of the protocluster (middle) and in the filament extending south-east of the protocluster (bottom). The blue and red lines show the observed \amm\ (1,1) and (2,2) spectra, respectively. The \amm\ (1,1) emission lines are offset by 1~K for clarity. The black lines show the model fits.} 
    \label{fig:spectra}
\end{figure}

Examples of \amm\ (1,1) and (2,2) emission lines in the protocluster and filaments are shown in Figure \ref{fig:spectra}.

\begin{figure*}
    \includegraphics[width=0.42\textwidth]{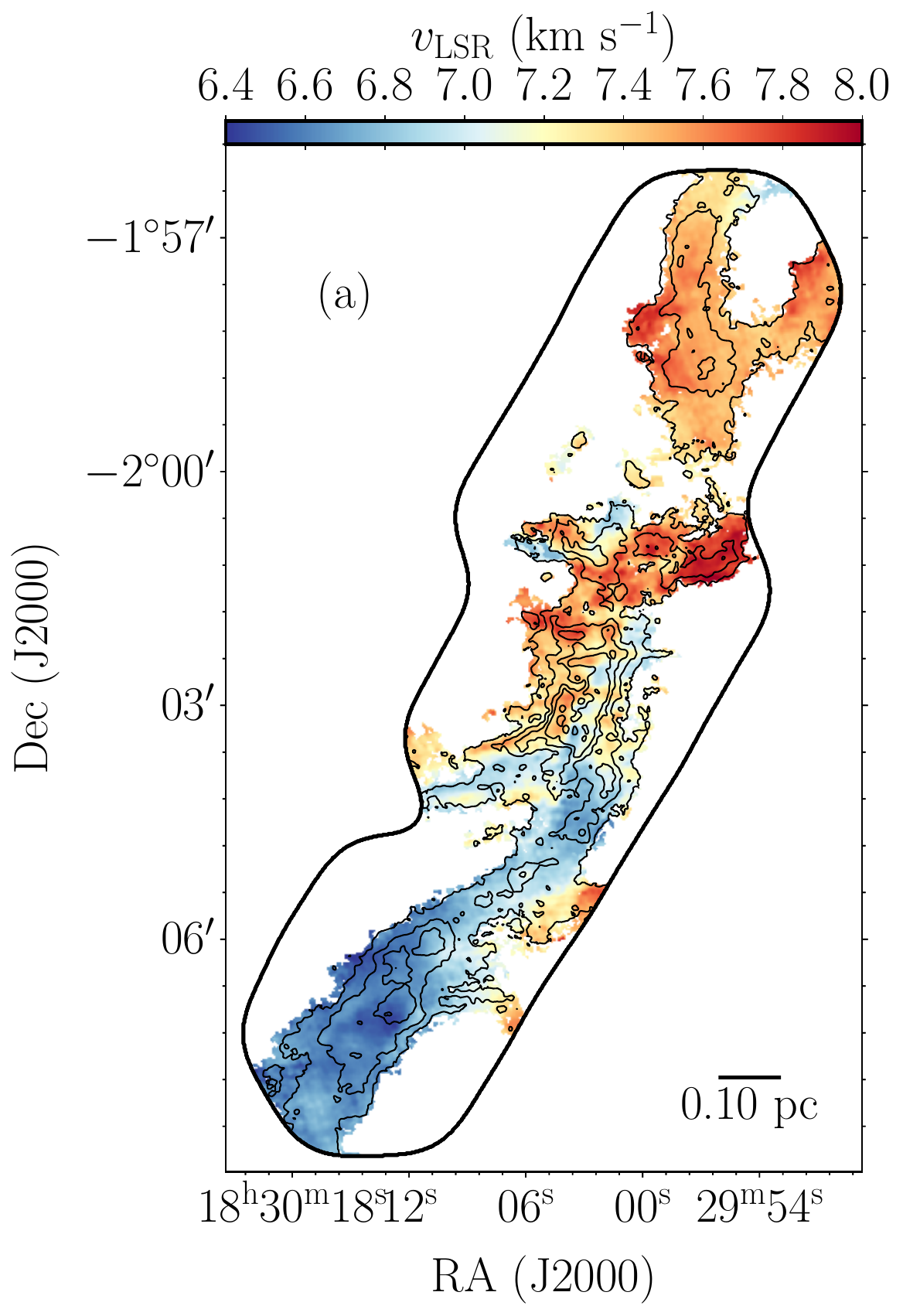}
    \includegraphics[width=0.41\textwidth]{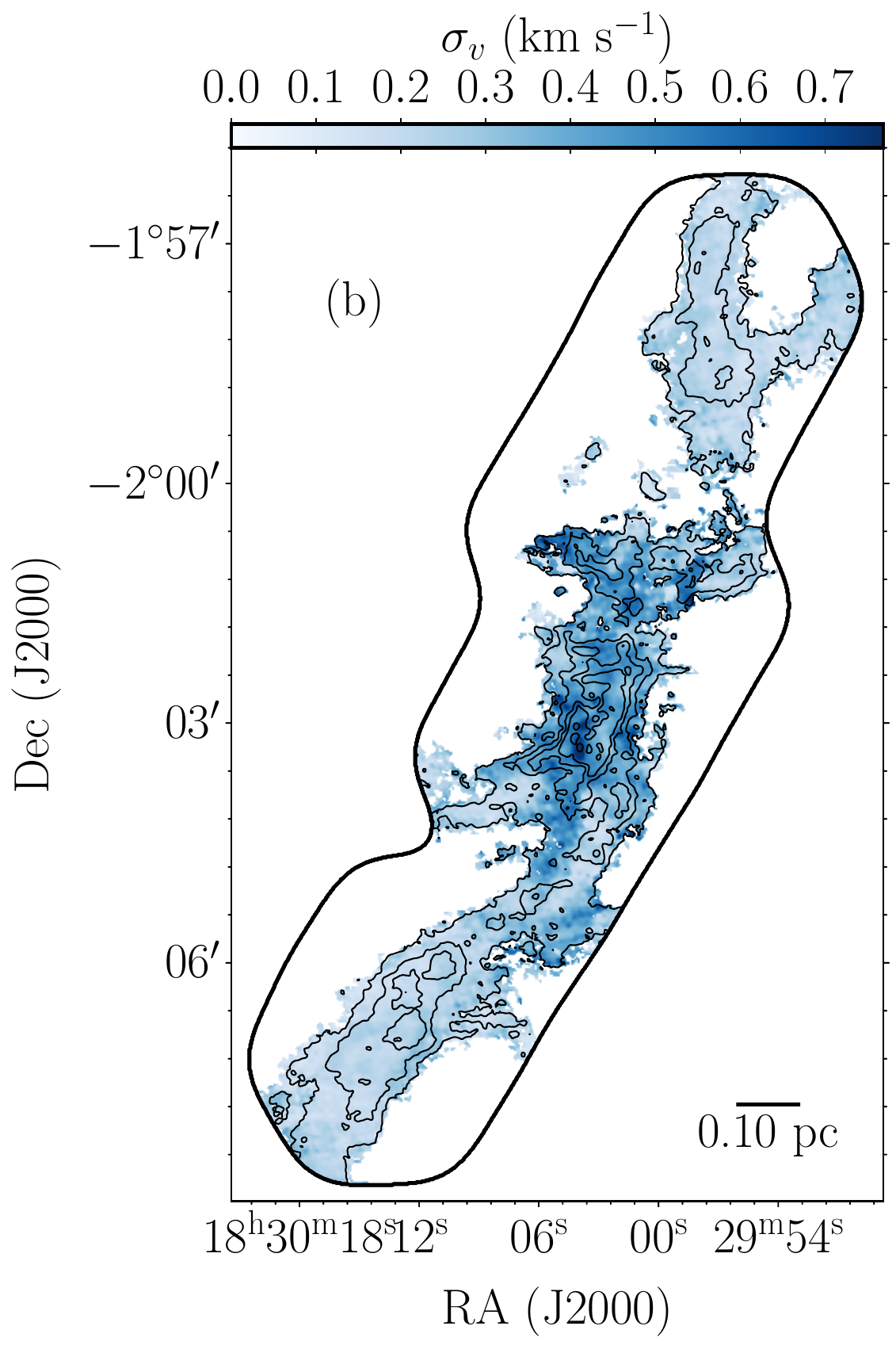} \\
    \includegraphics[width=0.42\textwidth]{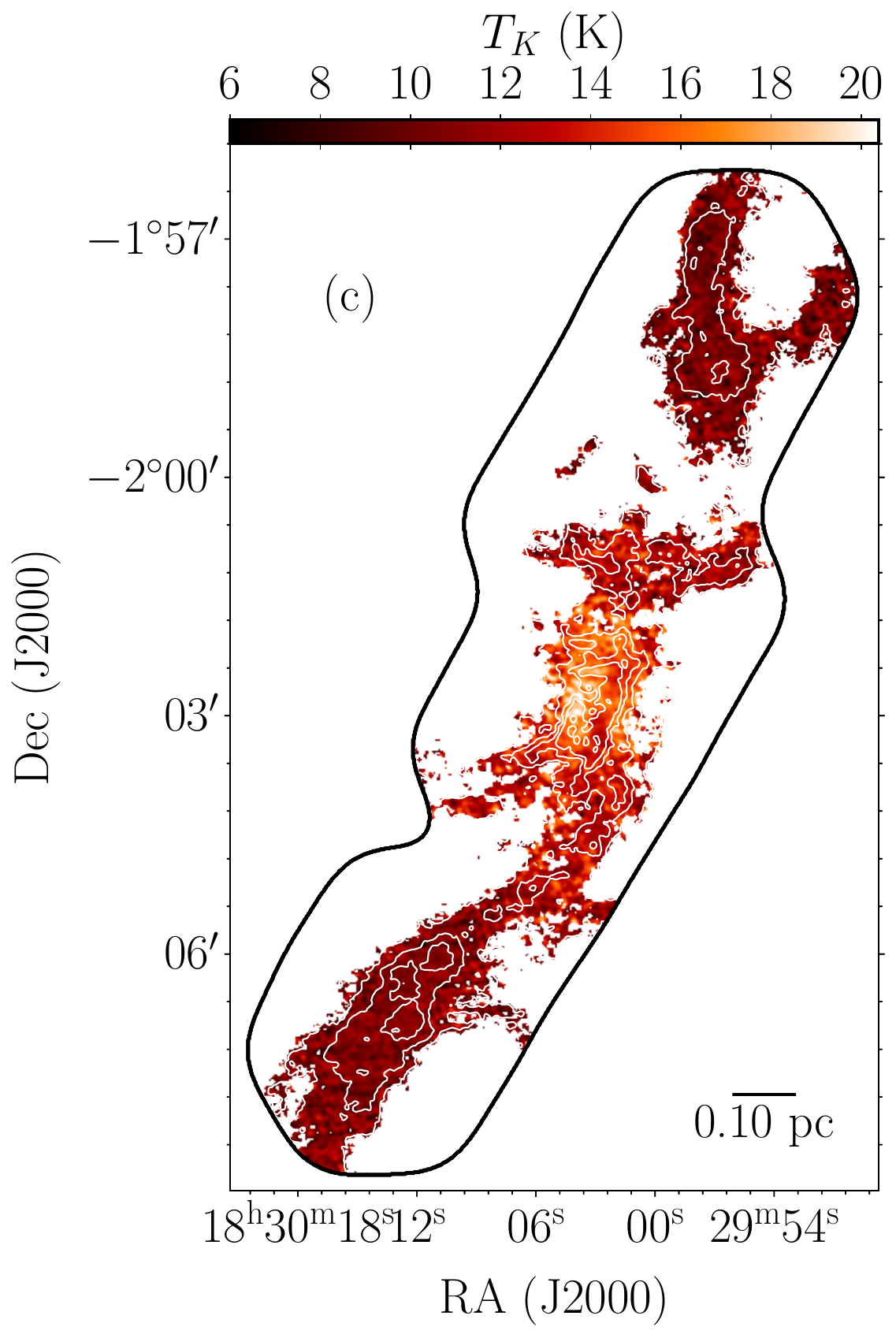}
    \includegraphics[width=0.44\textwidth]{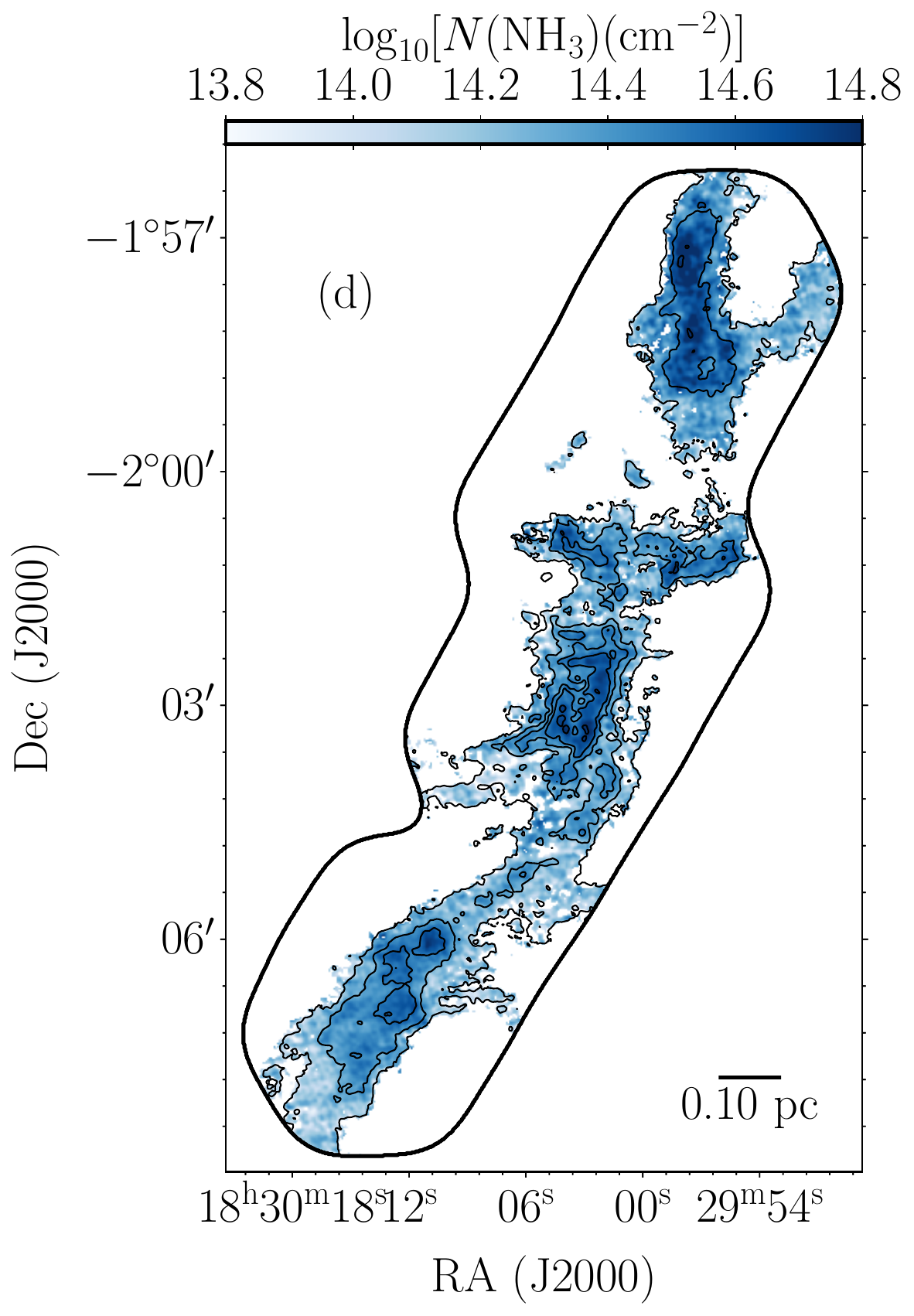}    
    \caption{Parameter maps obtained from the \amm(1,1) and \amm(2,2) line fitting. In all four plots, overlaid contours are of the \amm\ (1,1) integrated intensity data and begin at 1.8~K~\kms\ and increase in 1.8~K~\kms\ intervals. Top left: local standard of rest velocity towards Serpens South. Top right: total line width across Serpens South. Bottom left: map of kinetic gas temperature across Serpens South. Bottom right: map of \amm\ column density across Serpens South. }\label{fig:parameter_maps}
\end{figure*}

\subsubsection{\amm\ parameter maps}

Figure \ref{fig:parameter_maps} depicts the maps of the parameters obtained from the \amm\ (1,1) and \amm\ (2,2) line fitting. We overlay the \amm\ (1,1) integrated intensity contours on each of the maps with contours starting at 1.8 K \kms\ and increasing in 1.8 K \kms\ increments. 

Figure \ref{fig:parameter_maps} (top left) shows the local standard of rest velocity, \vlsr\, towards Serpens South which ranges from $\sim 6$--8~\kms\ across the map. \vlsr\ is lowest in the filament towards the southeast of the central cluster with values generally in the range of $\sim 6$--7~\kms, increasing in the northwestern filament up to values $\sim 7$--8~\kms. The implications of this line of sight velocity gradient across Serpens South will be discussed further below. This range of \amm\ \vlsr\ is consistent with \citet{friesen_2016} where Serpens South is traced with \amm\ at lower resolution and with \citet{fernandez_2014} where Serpens South is traced with N$_2$H$^{+}$ at a resolution of 7\arcsec, which is a similar resolution to our data.

Figure \ref{fig:parameter_maps} (top right) shows that the total velocity dispersion, \sigv\, towards Serpens South ranges from 0.2--0.6~\kms. We find that \sigv\ is generally broader towards the central cluster. Within the central cluster, the \sigv\ values range from about 0.4--0.6~\kms\ and within the filaments the values are within the approximate range of 0.1--0.4~\kms. This is comparable to the lower resolution data from \citet{friesen_2016} where they find that $\Delta v \sim 0.5$~\kms\ (FWHM) in the filaments, and $\sim 1-1.5$~\kms\ toward the cluster.

Across Serpens South, we see gas temperatures, $T_\mathrm{K}$, ranging from 8--24~K as shown in the  Figure \ref{fig:parameter_maps} (bottom left). We find that the gas temperature is greatest towards the central cluster of Serpens South, and is lower in the connecting filaments. The gas temperatures in the protocluster range from 15--24~K while the temperatures in the filaments range from 10--14~K. The mean gas temperature across all of Serpens South is 12~K with a standard deviation of 2~K. This mean gas temperature is slightly higher than the value of 11 K which was found in \citet{friesen_2016} with lower resolution data (32\arcsec) and also has a higher standard deviation across Serpens South (2~K versus 1~K). Using \amm\ data at a similar resolution (4\arcsec), \citet{dhabal_2019} find a similar median kinetic temperature of 12.4~K in NGC 1333, another nearby, clustered low-to-intermediate mass star forming region in Perseus. These gas temperatures are lower than those in dense molecular gas within high-mass star forming regions such as Orion A \citep{monsch_2018}. 

In general we find that the integrated intensity contours closely follow the \amm\ column density, \namm, with greater line emission corresponding to higher column density in the central cluster and the middle of the filametary regions as seen in Figure \ref{fig:parameter_maps} (bottom right). The molecular gas across Serpens South has a mean \namm\ $= 2.5\times10^{14}$~cm$^{-2}$ with a standard deviation of $1.2\times10^{14}$~cm$^{-2}$. The column density also has a total range of $4.8\times10^{13}$~cm$^{-2}$ to $1.1\times10^{15}$~cm$^{-2}$.

We show in Figure \ref{fig:nh3_33_fits} maps of the \amm\ (3,3) \vlsr\ and \sigv\ based on the single Gaussian fits, focusing on the cluster centre where \amm\ (3,3) is detected.  

\section{Analysis}\label{section:Analysis}

We next investigate the dense gas properties traced by the \amm\ observations. We first discuss the identification and stability of dense molecular cores within the filaments and cluster gas. We then show how the cores are connected to the large-scale gas structures, including their temperatures and gas motions.  

\subsection{\amm\ core identification and analysis}

\subsubsection{\amm\ core identification via dendrograms}
\label{sec:astrodendro}

Ammonia emission highlights cold, dense gas in molecular clouds \cite[e.g.][]{myers_benson}. 
In Serpens South, the dense gas traced by \amm\ shows small-scale features embedded within the larger-scale filaments and cluster region.
Hierarchical algorithms such as dendrograms decompose datasets into coherent structures. Using the astrodendro\footnote{\url{http://www.dendrograms.org}} package, we performed a hierarchical analysis of the Serpens South \amm\ (1,1) integrated intensity map to obtain the hierarchical nature of the cloud structure. 
Dendrograms work well in PPV space, but because the \amm\ (1,1) line is composed of 18 separate hyperfine components, the algorithm cannot distinguish between true separate components in velocity space and hyperfine line components. 
For this reason, we perform the dendrogram analysis on the \amm\ (1,1) integrated intensity map, as has been done toward other star forming regions \citep[e.g.,][]{keown_2019}. 
This also improves the SNR of the detected sources, with the caveat that any cores that overlap in the plane-of-the-sky but are separated in velocity space will be identified as single objects. 
A visual inspection shows, however, that apart from very near the cluster centre, the \amm\ lines are well-fit by a single velocity component.  

The dendrogram method works by identifying emission peaks in a map or cube above a certain threshold. The algorithm tracks the brightness of identified structures until they merge with adjacent structures. The top-level structures (``leaves") are connected to ``branches" which are then connected to the lowest level structures (``roots"). A more detailed description of dendrogram analysis is discussed in \citet{rosolowsky_2008}.

Three parameters determine how structures are identified: \texttt{min\_npix}, \texttt{min\_value}, and \texttt{min\_delta}. \texttt{min\_npix} is the minimum number of pixels of a structure, which we set to the number of pixels within the VLA beam.  \texttt{min\_value} is the minimum brightness of the peaks. \edit1{\texttt{min\_delta} is the  minimum difference in brightness between individual structures before they are merged into one structure.} \texttt{min\_value} and \texttt{min\_delta} were set to integer multiples of the mean rms noise ($\sigma_\mathrm{rms}$ = 0.14~K~\kms), where \texttt{min\_value} $= 5 \times\sigma_\mathrm{rms}$ and \texttt{min\_delta} $= 3 \times\sigma_\mathrm{rms}$ \edit1{so that any leaf that is locally less than $3 \times\sigma_\mathrm{rms}$ tall will be combined with its neighbouring leaf}. These values removed noise sources from the dendrogram analysis without removing all of the low-brightness peaks in the emission. We note that the dendrogram algorithm is not sensitive to small changes in these parameters \citep{rosolowsky_2008}. Changing these parameters alters the number of leaves that are obtained from the dendrogram analysis but does not significantly change the overall results.

\begin{figure*}
    \centering
    \includegraphics[width=0.45\textwidth]{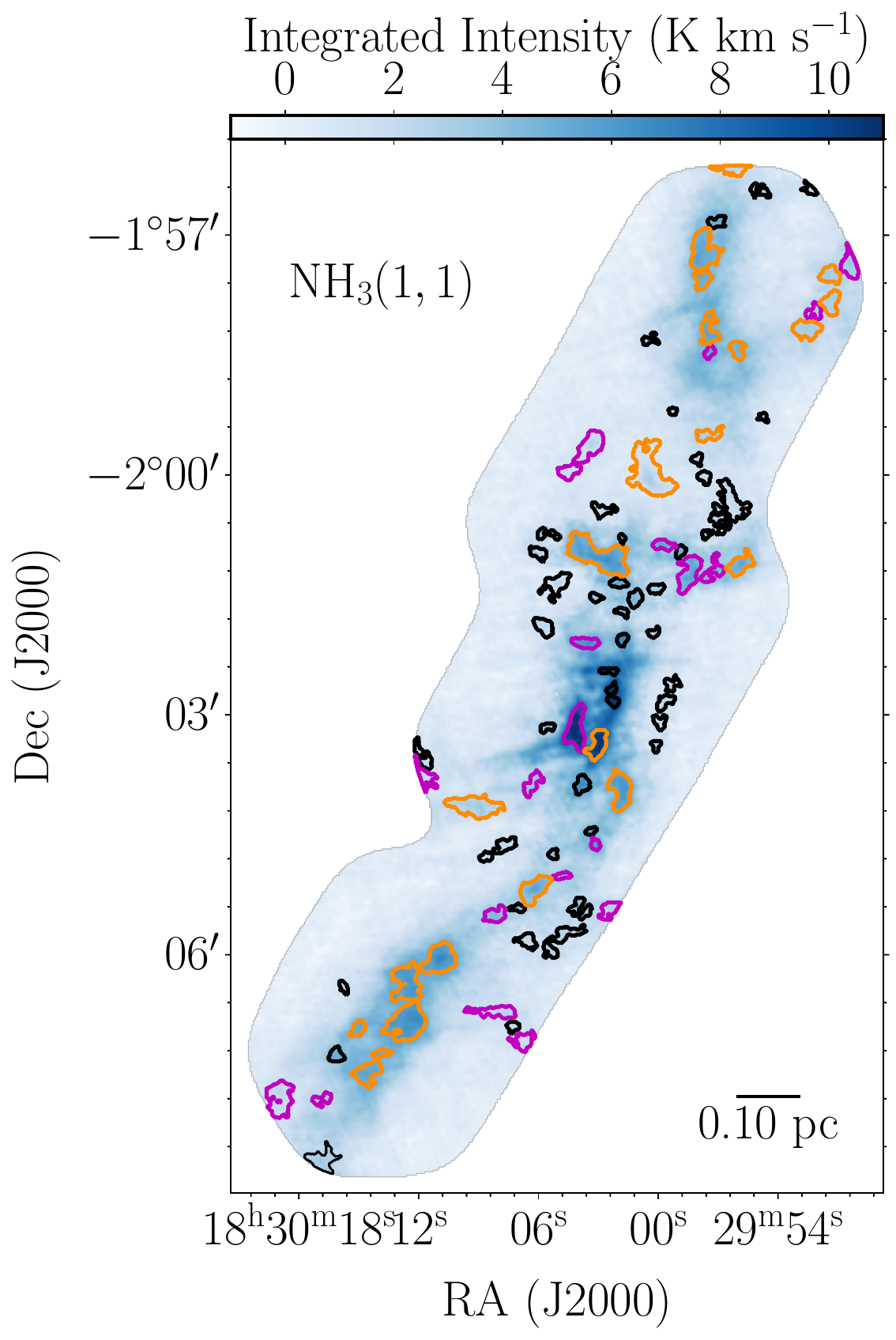}
    \includegraphics[width=0.45\textwidth]{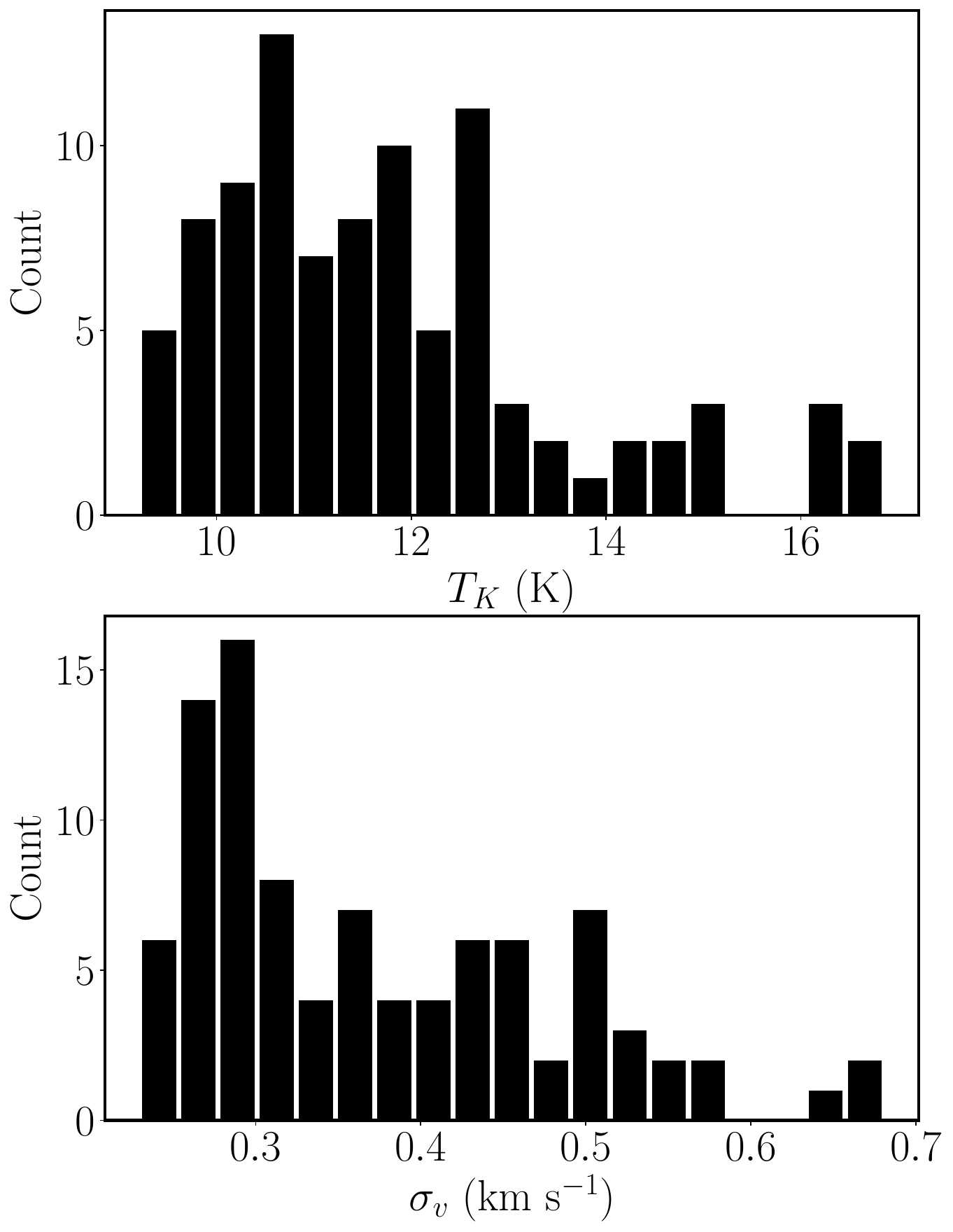}
    
    \caption{Left: \amm(1,1) integrated intensity map of Serpens South. Contours outline the 94 cores (leaves) identified from the dendrogram analysis that were used in the analysis of this paper. Black contours outline unbound cores with $\alpha>2$ following analysis in Section \ref{sec:virial}. Purple contours outline cores with $1<\alpha<2$. Orange contours outline bound cores with $\alpha<1$. Right: Distribution of gas temperature (top) and velocity dispersion (bottom) obtained from line fitting of the \amm\ emission of the 94 identified cores in Serpens South.}
    \label{fig:mom0cores}
\end{figure*}

With these three parameters determined, we obtain a dendrogram that contains 108 leaves and 107 branches. 
In this study, we focus primarily on the identified leaves, which correspond to dense cores of gas embedded within the larger filamentary structures of Serpens South.
We refer to the leaves as cores going forward. 
Figure \ref{fig:mom0cores} shows again the integrated \amm\ (1,1) intensity toward Serpens South, where the extents of the dendrogram leaves are outlined with coloured contours.

We then determine whether cores contain an embedded protostar (Class 0, Class I or flat-spectrum) using the catalog of protostars described in \ref{section:moment_maps}. We determine whether there is a protostar within a distance of the mean radius of all cores (0.01pc) from the center of each core. With this analysis we find that only 7 of the cores contain an embedded protostar. 

For each core, we create a mask corresponding to all pixels belonging to that core as determined by \textit{astrodendro}. The mask defines the exact area of the core, $A$. We define the effective radius of each structure as $R = \sqrt{A/\pi}$. 
We then apply this mask to each of the parameter maps in Figure \ref{fig:parameter_maps} and calculate the mean and standard deviation of each parameter for each core. 

When performing the line fitting, there are some pixels that contain NH$_3$ (1,1) emission, but no strongly detected NH$_3$ (2,2) emission. This means that although cores were identified with the NH$_3$ (1,1) integrated intensity, the maps of the gas temperature and column density do not have well determined values for some cores. There were 14 such cores and these cores are included in Table \ref{tab:core_properties} but are excluded from further analysis, leaving a total of 94 cores. Table \ref{tab:core_properties} contains the core locations, radii, and \amm\ fit parameters for all 108 cores identified with the dendrogram analysis. We additionally show in Figure \ref{fig:mom0cores} the histograms of mean $T_K$ and $\sigma_v$ across the 94 cores that are used in further analysis. 

\subsubsection{Core masses}\label{section:core_masses}

We calculate the masses of the cores directly from the \amm\ emission. 
To do so, we must determine an appropriate abundance of \amm\ to H$_2$, \xamm = \namm/\nh. 
Since (sub)millimeter continuum data, typically used to derive \nh, is not currently available at matching 5\arcsec\ resolution, we calculate \xamm\ across Serpens South from \namm\ and \nh\ column densities at 32\arcsec\ and 36\arcsec\ resolution, respectively \citep{friesen_2016,singh_2022}.
On these larger scales, we find that the mean \xamm\ $ = 1.5\times10^{-8}$ with a standard deviation of $0.5\times10^{-8}$. This value of \xamm\ is similar to that found in other star-forming regions, such as Orion and Cepheus \citep{harju_1993}. 
\edit1{The \amm\ abundance map on larger scales shows some systematic variation between the filaments and the cluster (see Figure 5 in \citeauthor{friesen_2016}).}
\edit1{Because of this variation in \xamm\ across Serpens South, we calculate an \xamm\ value for each core using the 32\arcsec\ resolution data. }
\edit1{While there may still be variations in \xamm\ on smaller scales not resolved by the lower resolution maps, this removes any systematic changes in the \amm\ abundance between the filaments and cluster gas, and thus in the following mass calculations. } 

The cores in Serpens South are often located along filaments or near the central cluster where \amm\ emission is extended. Thus, it is necessary to subtract the background from the column density before determining the core mass.  
To subtract background emission, we determined the mean background \namm\ value within an annulus around each core using Astropy's \textit{photutils} package \footnote{\url{https://photutils.readthedocs.io}}. We defined the annulus around each core with an inner radius of 2 times the effective core radius obtained by approximating each core as an ellipse and an outer radius of 3 times the effective core radius. This annulus size ensures that we remain close enough so that we are indeed sampling each cores' background and not so far that we are no longer near the core or overlapping nearby cores. We then obtained the mean value within each annulus and subtracted this value from each pixel within each dendrogram contour. Some cores further away from the filaments were unchanged from this background subtraction, because there is no \amm\ column density around those cores, but this is acceptable because these cores do not have significant background gas. 

With these assumptions, we calculate the core masses. We use two methods to estimate the masses from the \namm. Of the 94 cores, we find that 54 of them have \namm\ data in each pixel and an additional 16 cores have \namm\ data in at least 75\% of pixels. Where we have \namm\ values for at least 75\% of pixels in each core, we calculate the mass by summing all of the \namm:

\begin{equation}
M_\mathrm{core} = A_{\mathrm{pixel}} \times \frac{\sum N({\mathrm{NH}_{3}})} {X(\mathrm{NH}_3)} \times \mu m_\mathrm{H}
    \label{eqn:mass_sum}
\end{equation}

where $A_{\mathrm{pixel}}$ is the area of each pixel and $\sum N({\mathrm{NH}_{3}})$ is the sum of all \namm\ in each core, $m_{\mathrm{H}}$ is the molecular mass of hydrogen and $\mu=2.72$ \citep{kauffmann_2008}. For the 16 cores with \namm\ in at least 75\% of pixels, but less than 100\% of pixels, we might be underestimating some of the core masses following this method.

If we take the minimum \namm\ pixel value within each of these cores and assume this is the \namm\ value for the missing pixels, we will underestimate the mass by at most 6\% for 9 of these cores, and by between 10\% -- 22\% for the other 5 cores.

For the 24 cores that do not have \namm\ data in at least 75\% of pixels, we calculate the core mass using the mean \namm. These core masses were calculated as: 

\begin{equation}
M_\mathrm{core} = A_{\mathrm{exact}} \times \frac{N({\mathrm{NH}_{3}})_{\mathrm{mean}}} {X(\mathrm{ NH}_3)} \times \mu m_\mathrm{H}
    \label{eqn:mass_mean}
\end{equation}

where $A_{\mathrm{exact}}$ is the exact area of each core (obtained from the dendrogram analysis), and \namm$_{\mathrm{mean}}$ is the mean column density across each core. 
Overall, we find that the group of cores that are calculated by summing the \namm\ rather than taking the mean have masses that are greater by $\sim$ 30\% when comparing the cores calculated via each method.

Across all cores, we find a \edit1{mean core mass of 1.7 \msun\  and median of 0.8~\msun}. 

Without any background subtraction, the mean core mass of the cores would be slightly greater, \edit1{$\langle M \rangle \sim 2$~\msun}. 
This can be thought of as an upper limit of the core masses because this mass calculation sums all \amm\ emission along the line of sight. 

We find that the \edit1{mass-radius relationship of the cores is best fit with a power law index of 2.27~$\pm$~0.11.} The fit was obtained using a nonlinear least squares fitting method with \texttt{curve\_fit} from the \texttt{scipy} package, and the uncertainty in the parameter was taken to be the square root of the variance. This value is between a value of two, which is expected for cores with constant surface density and three, which is expected for cores of constant volume density.

\subsubsection{Virial analysis of \amm\ cores}
\label{sec:virial}

 We use the virial theorem to assess the stability of dense cores. If a core has a mass greater than the virial mass, it should be actively collapsing unless there is another form of support against gravity. When only considering kinetic and gravitational potential energy terms in the virial equation, the virial mass is defined as  \citep{bertoldi_1992}

\begin{equation}
M_\mathrm{vir} = \frac{5\sigma^2R}{aG}
    \label{eqn: Virial Mass}
\end{equation}

\noindent where $R$ is the radius of the cores, $G$ is the gravitational constant and $\sigma$ is the velocity dispersion and $a$ is a correction factor that depends on the density profile of the core.
We set $a=1$ because this assumption is consistent with low-mass cores \citep{singh_2021}.
\edit1{Because the uncertainties in $\sigma$ from our line fits are small, our assumption for $a$ likely dominates the uncertainty in this calculation.}

We calculate the velocity dispersion as in \citet{bertoldi_1992}:

\begin{equation}
\sigma^2 = \sigma_V^2+\left<\sigma_{\mathrm{LS}}\right>^2-\frac{k_\mathrm{B}T}{m_{\mathrm{NH}_3}}+\frac{k_BT}{\mu m_{\mathrm{H}}}
    \label{eqn: sigma}
\end{equation}

\edit1{\noindent where $T$ is the mean kinetic temperature over the core, $k_B$ is the Boltzmann constant, $\sigma_V$ is the dispersion of the mean \vlsr\ for each core, $\left<\sigma_{LS}\right>$ is the measured mean line width over the face of each core, $m_{\mbox{\tiny{NH$_3$}}}$ is the molecular mass of ammonia, $m_{\mbox{\tiny{H}}}$ is the mass of hydrogen, and $\mu = 2.37$ for molecular gas \citep{kauffmann_2008}. The second term in this equation takes into account the bulk motion in the cores, stemming from variations of the line-of-sight velocity corresponding to internal motions. As discussed by \citet{singh_2021}, this bulk motion term can contribute significantly to the kinetic energy of dense structures. If not taken into account, calculated virial parameters may be much lower than the true values. The term is small for most cores in this study. Across all cores, this term has a mean of 0.05~\kms\ and for the largest cores, this term is up to $\sim 0.10$~\kms. }

\begin{figure}
    \centering
    \includegraphics[width=0.45\textwidth]{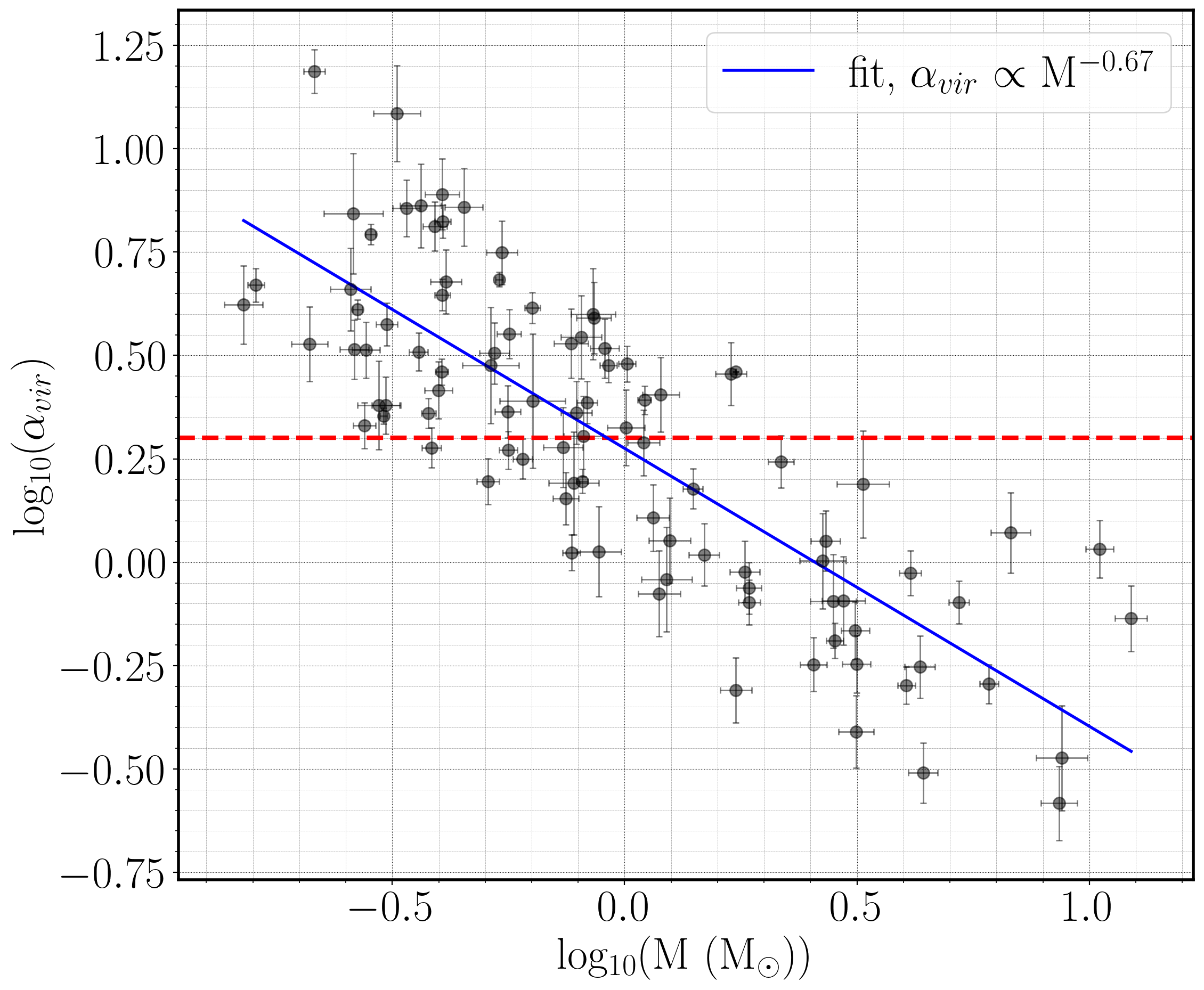}
    \caption{The relationship between mass and virial parameter for cores in Serpens South. The blue line shows the best fit of the data with a \edit1{power law index of  $-0.67 \pm 0.05$ . The red dashed line indicates a virial parameter of 2. Of the 94 cores, 51 of them have a virial parameter greater than 2, indicating that they are stable against their self-gravity.}
    \label{fig:mass_virial_param}}
\end{figure}

To assess the stability of the cores, we then calculate the virial parameter as $\alpha = M_\mathrm{vir}/M_\mathrm{core}$.
Across all cores, we find a \edit1{median virial parameter of 2.3 and a mean virial parameter of 2.8. The virial parameters vary between 0.3 and 15.4. We find that most cores (54\%) have virial parameters greater than the critical value of $\alpha=2$, indicating that they are not gravitationally bound. Only 24\% of the cores are sub-virial ($\alpha<1$)}. \edit1{Core virial parameters are plotted against core masses in Figure \ref{fig:mass_virial_param}.} We discuss the implications of these core stability results, and their correlation with location within Serpens South, in section \ref{section:stability_discussion}. 

\subsection{Large-scale properties of the dense gas}
\label{sec:kinematics}

\begin{figure}
    \centering
    \includegraphics[trim={130 70 110 60}, clip,width=0.49\textwidth]{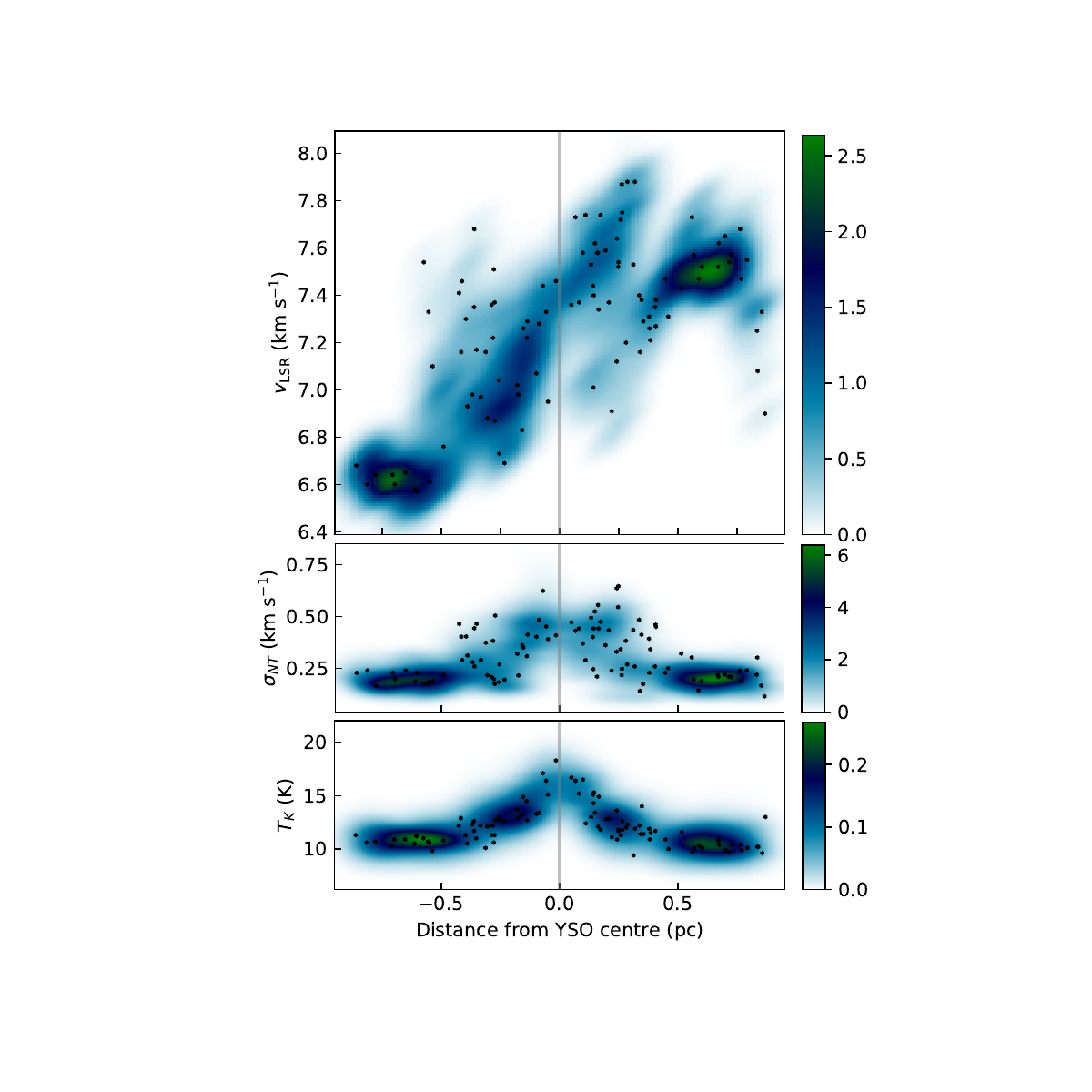}
    \caption{Kernel Distribution Estimation (KDE) plot of the \vlsr\ (top), \signt\ (middle), and $T_K$ (bottom) as a function of distance from the YSO cluster centre. The colour scale shows the density of points. The \vlsr\ shows a clear velocity gradient through the cluster, while \sigv\ and $T_K$ increase significantly within $\sim 0.25$~pc of the centre. In all plots, black points show the mean values for individual \amm\ cores.}
    \label{fig:vlsr_kde}
\end{figure}

Here, we examine the gas kinematics and temperature with respect to the YSO cluster. We define the cluster centre by averaging the positions of the two stellar groups near the centre of Serpens South obtained from \citet{sun_2022b}, as these two groups have a similar number of stars. The cluster centre has an RA of 18:30:3.866 and a Dec of -2:2:37.521.

The line-of-sight velocity in Figure \ref{fig:parameter_maps} shows a clear velocity gradient 
\edit1{on large scales, where the southern filament is significantly blue-shifted and the northern filament significantly red-shifted relative to the gas near the central cluster.}
The velocity gradient along the southern filament has been previously identified at lower resolution than presented here \citep{kirk_2013,tanaka_2013,fernandez_2014, friesen_2016}, and is generally interpreted as arising from mass flow toward the central cluster, driving accretion. 

We show in Figure \ref{fig:vlsr_kde} (top) the kernel density estimations (KDEs) of the \amm\ \vlsr\ as a function of radial distance from the cluster centre. Because we have primarily observed the north and south filaments in Serpens South, the radial distance is generally a good estimate of \edit1{(projected)} distance along the filament from the cluster center. We have overlaid on the Figure the locations and mean \vlsr\ values of the \amm\ cores identified in \ref{sec:astrodendro}. 

\edit1{We find a velocity gradient of $0.70 \pm 0.25$~km~s$^{-1}$~pc$^{-1}$ through a linear fit of the \vlsr\ toward the southern filament to a distance of 0.5~pc from the cluster center.}
Using 40\arcsec\ data, \citet{kirk_2013} find a velocity gradient in Serpens South southern filament of $\sim 0.7 \pm 0.1$~\kmspc\ (converting their result, assuming $d = 260$~pc, to a $d = 440$~pc), in agreement with our results within the uncertainties. 
\edit1{By analysing the gas energetics, \citet{kirk_2013} argued that the southern velocity gradient is due to mass flow along the filament rather than large-scale rotation. }
\edit1{We can make similar arguments that the velocity gradient to the north also represents flow toward the cluster centre along a similar inclination as in the southern filament, such that the expected velocity gradient should be red-shifted relative to the cluster gas as viewed from our perspective.}
\edit1{A linear fit to all \vlsr\ data within $\sim 0.35$~pc of the cluster center does not produce a significant result, however, despite the clear transition between blue- and red-shifted emission at the cluster location visible in Figure \ref{fig:parameter_maps}. }
\edit1{On smaller scales, the velocity gradient near the cluster shows a change in orientation, such that there is both red- and blue-shifted gas within the fit region resulting in a poor fit with a simple linear distance to the cluster center. On these scales the observed motions could be a result of both inflow and rotation.}

This \edit1{suggests} that mass accretion onto the cluster \edit1{could be} proceeding both from the southern and northern filaments, \edit1{although the kinematics in the north are more complex than can be explained solely by inflow}.
The velocity gradient\edit1{s do} not extend the full length of our observations, however, and gas at the southern and northern ends of the filaments show a flattening or offset in velocity compared to the regions closer to the cluster. These regions of offset \vlsr\ correspond to the larger \amm\ integrated intensity structures seen in the filaments to the north and south of the central cluster in Figure \ref{fig:protostars}.  

We additionally show in Figure \ref{fig:vlsr_kde} the KDE of the non-thermal velocity dispersion (middle panel) and the gas kinetic temperature, \tkin\ (bottom panel), as a function of distance from the YSO cluster. The non-thermal velocity dispersion is calculated from the observed dispersion $\sigma_{v}$ via
\begin{equation}
    \sigma_{NT}^2 = \sigma_{v}^2 - \frac{k_B T_K}{m_\mathrm{NH_3}}
\end{equation}
where $T_K$ is the \amm-derived gas temperature, thus removing any contribution to the velocity dispersion driven by changes in gas temperature. We are able to measure \signt\ where we have reliable \tkin\ measurements only. Over much of the observed region, the non-thermal velocity dispersion values, with a median $\sigma_\mathrm{NT} \sim 0.23$~\kms, remain slightly greater than the sound speed at the mean $T_K \sim 12$~K, $c_s = 0.20$~\kms. 

\begin{figure*}
    \includegraphics[trim = 20 50 40 200, clip, width=0.8\textwidth]{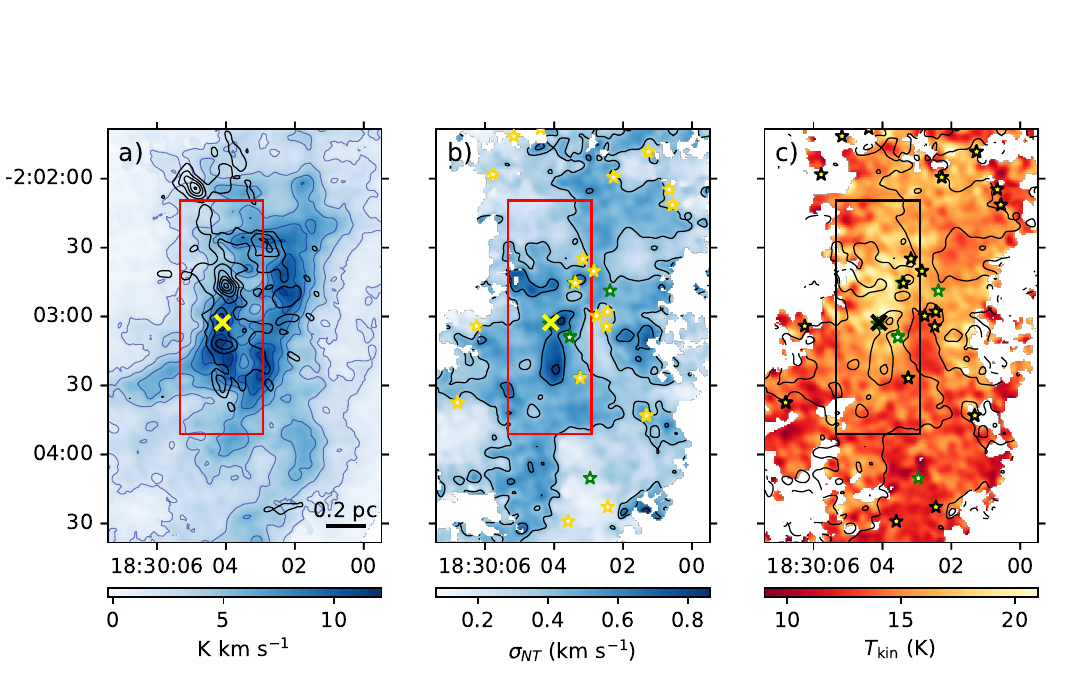}
    \includegraphics[trim = 20 0 20 20, clip, width=0.19\textwidth]{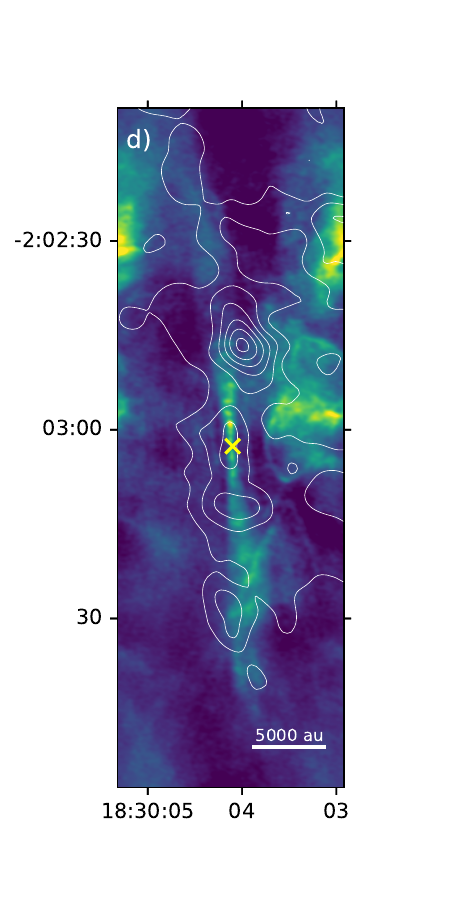}
    \caption{\edit1{a) Integrated \amm\ (1,1) intensity (colormap and blue contours) overlaid with \amm\ (3,3) integrated intensity contours (black) within the black box in Figure \ref{fig:cont_33}. In all panels, the Class 0 driving source of the collimated outflow, CARMA-7, is indicated by the `x' \citep{plunkett_2015_outflows}. In a) through c) the inset box highlights the area shown in panel d). b) Non-thermal \amm\ velocity dispersion \signt\ in the same region as a). Black contours show \signt\ $= 0.4, 0.6, 0.8$~\kms, or roughly 2, 4, and 6 times the thermal sound speed $c_s$ in the cluster center. c) Gas temperature \tkin\ (color) with \signt\ contours as in b). d) CO 2-1 intensity at 0\farcs9 $\times$ 0\farcs6 resolution \citep[color;][]{plunkett_2015_epi}, overlaid with \amm\ (3,3) integrated intensity contours as in a). Where the data are available to compare, increased \signt, \tkin, and \amm\ (3,3) emission all align with the collimated outflow. \label{fig:outflow}}}
\end{figure*}

Both \signt\ and \tkin\ increase significantly toward the cluster centre. 
Figure \ref{fig:vlsr_kde} shows that the gas temperature shows a smooth increase at decreasing distance from the YSO centre, \edit1{while} \signt\ appears to have a sharper jump from lower values at small distances ($\lesssim 0.3$~pc), with larger scatter than \tkin. 
\edit1{To better show the increase in \signt\ and \tkin\ spatially near the protostars, we present zoomed-in maps of the \amm\ emission, \signt, and \tkin, respectively, toward the central protostellar cluster in Figure \ref{fig:outflow} a) - c).}
\edit1{The non-thermal velocity dispersions are supersonic throughout the cluster region, with the highest values found along the direction of the collimated outflow driven by the Class 0/I source CARMA-7 (highlighted on the Figure). Similarly, warmer gas is co-located with CARMA-7 and the surrounding small groups of protostars, with \tkin\ values greater in general where \signt\ is also increased.}

The increased \signt\ values toward the cluster center could be driven by several mechanisms, including protostellar feedback and mass accretion, \edit1{and we discuss this} further in \ref{section:cluster_effect}. 

\begin{figure}
    \includegraphics[width=0.49\textwidth,trim={0 0 0 0},clip]{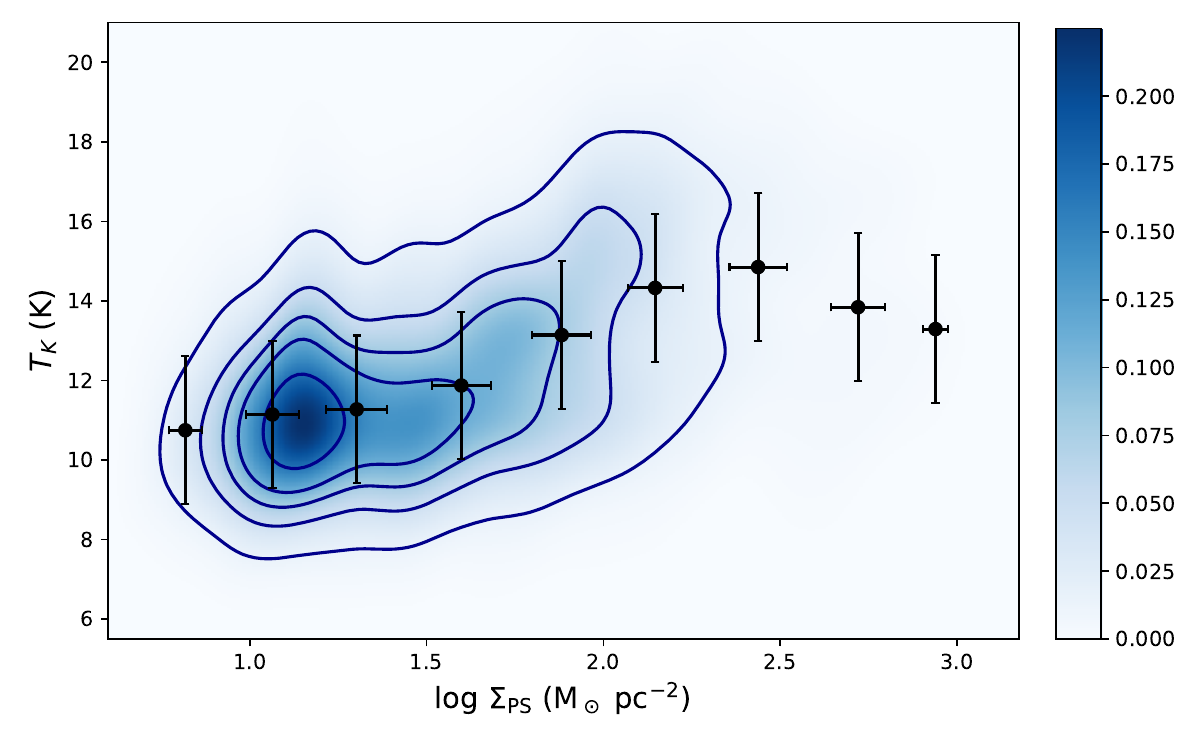}
    \includegraphics[width=0.48\textwidth,trim={0 0 0 7},clip]{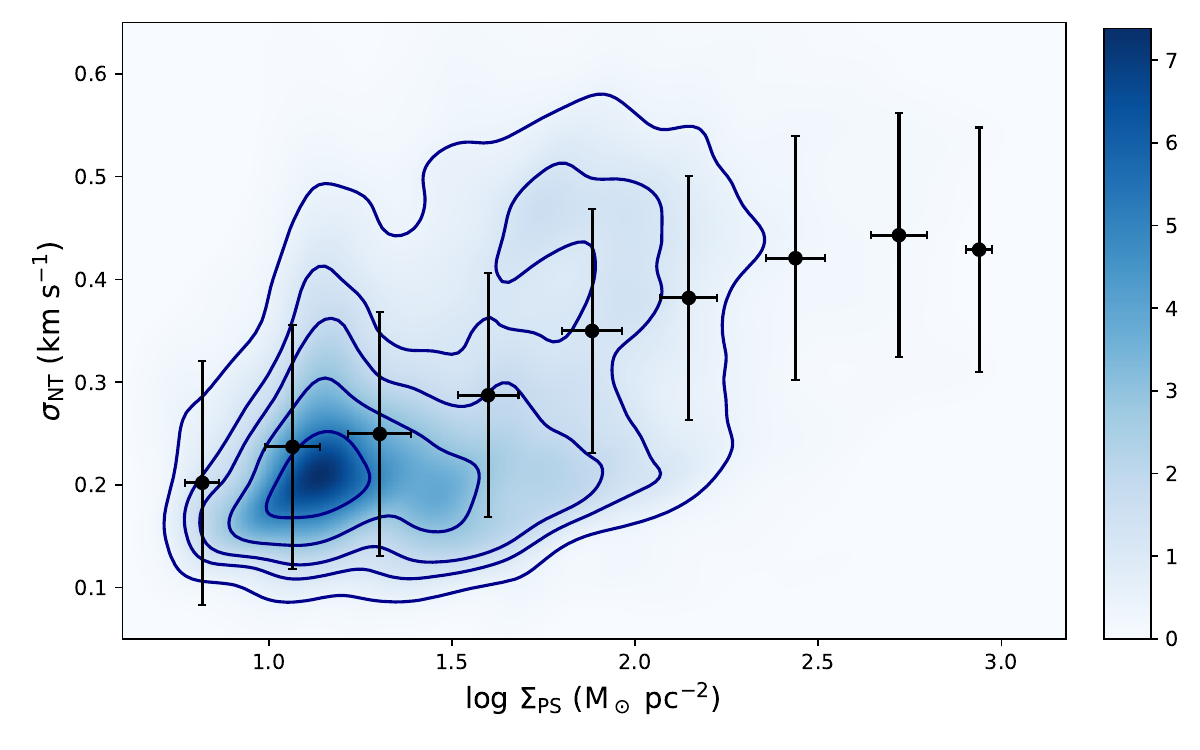}
    \includegraphics[width=0.49\textwidth]{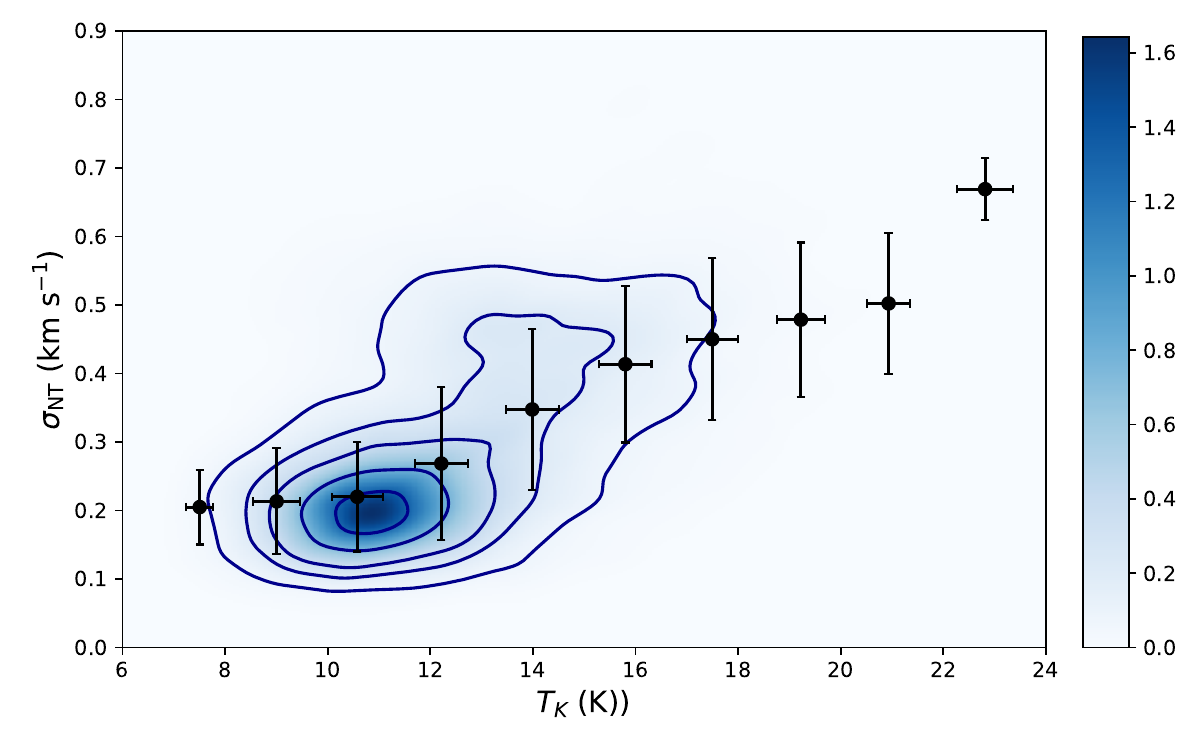}
    \caption{Top and middle panels: Kernel density estimates (KDEs) of the distribution of \tkin\ (top) and \signt\ (middle) with protostellar surface density \sigps. Contours represent density levels of 0.1 through 0.9 in intervals of 0.2. Black data points show binned results with 1-$\sigma$ bars. Bottom: KDE of the distribution of \signt\ with \tkin, with contours and data points as in the top two panels. \label{fig:psd}}
\end{figure}

\subsubsection{Protostellar surface density}

\edit1{Both the gas temperature \tkin\ and the non-thermal velocity dispersion \signt\ increase toward the YSO cluster center.} We next examine the gas properties with respect to the protostellar surface density, \sigps. We calculate \sigps\ following
\begin{equation}
    \Sigma_\mathrm{PS} = \frac{n-1}{\pi r_n^2} \times \langle M_* \rangle
\end{equation}
where $r_n$ is the projected separation between a protostar and its $n^{th}$ nearest neighbor \citep{kirk_2011}, which we convert to parsecs given the distance to Serpens South. We adopt $n = 4$, reducing the resulting uncertainty in \sigps\ while broadly retaining spatial resolution  \citep{li_2019_psd}. We merge the protostellar catalogs from \citet{sun_2022} and \citet{pokhrel_2023}, and include identified Class 0, I and flat spectrum protostars. We further assume a mean protostellar mass $\langle M_* \rangle = 0.5$~M$_\odot$ \citep{evans_2009} to calculate \sigps\ in units of M$_\odot$~pc$^{-2}$.  

We show the resulting correlations between \sigps, \tkin, and \signt\ in Figure \ref{fig:psd} as KDEs, with data points overlaid showing the mean and standard deviation of each parameter in a set of bins. Because \sigps\ is greatest at the cluster center, we see that both \tkin\ and \signt\ increase with increasing protostellar surface density, as expected from Figure \ref{fig:parameter_maps}. The correlation is strongest, however, between \tkin\ and \signt\ themselves: dense gas that is warmer in Serpens South also shows greater non-thermal velocity dispersions. The strong correlation between the two parameters implies that both the higher temperatures and greater non-thermal velocity dispersions are driven by the same mechanism. In the next section, we argue that this mechanism is feedback from the central stellar cluster. 

\section{Discussion}\label{section:discussion}

\subsection{Stability of cores in Serpens South}\label{section:stability_discussion}

The sub-virial cores are bound and should be actively collapsing, unless there is another form of pressure support against gravity, such as could be provided by magnetic fields. The super-virial cores, on the other hand, are unbound by gravity, but could be confined by other forms of external pressure support. Other studies find that many cores are not bound by gravity alone, but require other forms of pressure confinement. Using lower resolution (32\arcsec) data, \citet{kirk_2017} argue that many cores in Orion A are confined by pressure from the weight of the ambient cloud material. \citet{kerr_2019} argue that many small cores in the nearby star forming regions L1688 in Ophiuchus, NGC 1333 in Perseus, and B18 in Taurus are confined by external turbulent pressure.

Figure \ref{fig:mass_virial_param} shows the relation between mass and virial parameter for the cores. There is a strong trend of decreasing virial parameter with increasing core mass with a power law index of \edit1{-0.67~$\pm$~0.05}. This trend is seen in other regions of star formation as well \edit1{\citep[e.g.,][find power law indices of $-0.50\pm0.10$, $-0.67\pm0.24$, $-0.68\pm0.03$ and $-0.54\pm0.08$ for the Rosette, Orion~B Ophiunchus, and Cepheus clumps respectively]{bertoldi_1992}}. The line of best fit in Figure \ref{fig:mass_virial_param} crosses $\alpha=2$ at a mass of \edit1{0.92~\msun\ and $\alpha=1$ at a mass of 2.57~\msun.} These points mark the transition from super-virial cores to sub-virial cores.   

\edit1{There are several factors that could affect this trend and change the stability of cores.}
First, if our background subtraction method were too conservative, these core masses would be underestimated, and the virial parameters overestimated. As described in Section \ref{section:core_masses}, our background subtraction method gives mass values in between those obtained by subtracting the minimum \namm\ from the core and those with no background subtraction. \edit1{Second, our estimation of \xamm\ is obtained from lower resolution data that does not match the resolution of the \namm\ data. Unresolved abundance variations could slightly alter the calculation of the core masses and virial parameters, but this is unlikely to be a systematic effect. }

\begin{figure}
    \centering
    \includegraphics[width=0.44\textwidth]{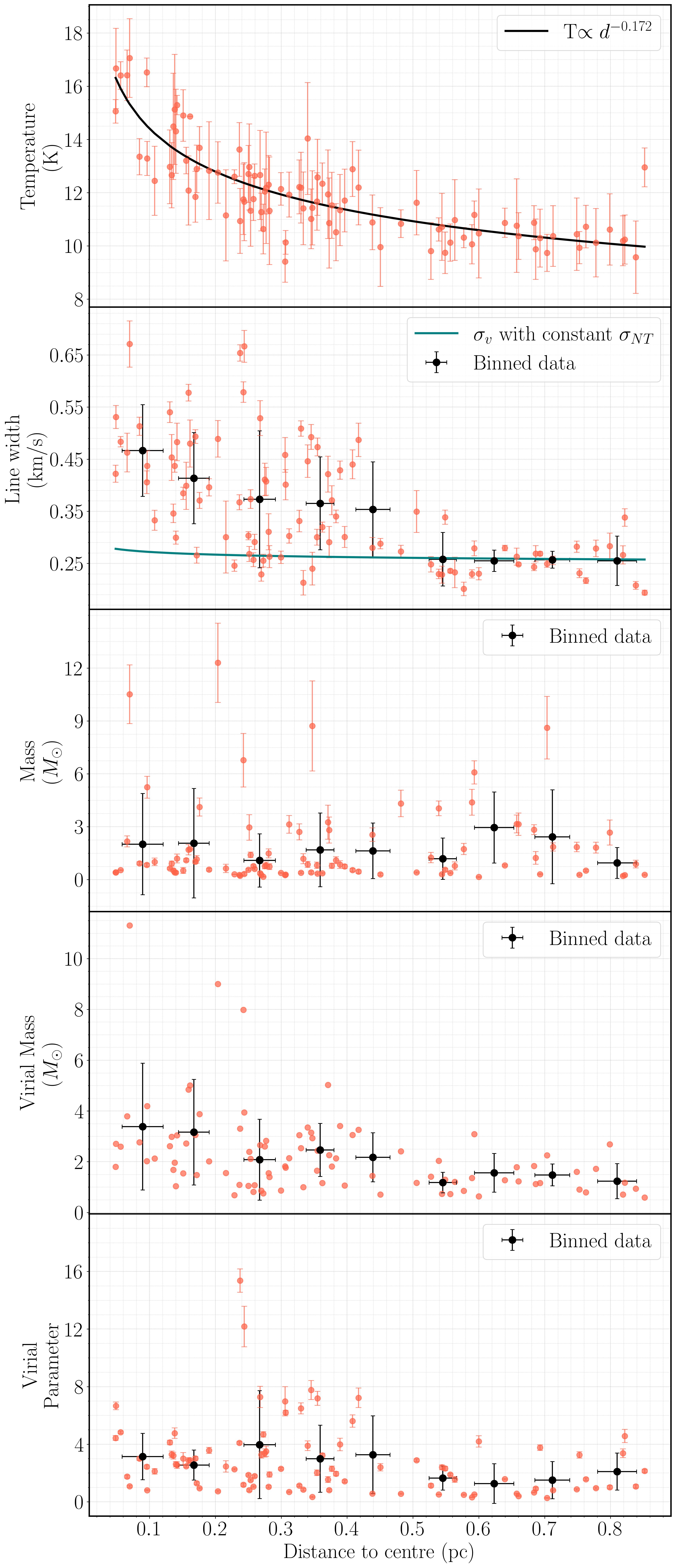}
    \caption{The relationship between various parameters to the distance to the centre of the Serpens South protocluster. Top: the temperature is best fit with a power law index of $-0.17 \pm 0.01$ (black line). For the plots of line width, virial mass and virial parameter, the black points are the data binned with bins of width 0.1~pc with errorbars that indicate the standard deviation of points within each bin. \label{fig:parameters_vs_distance_to_centre}}
\end{figure}

In Figure \ref{fig:mom0cores}, we use different coloured outlines to depict the stability of each core. Black contours outline cores with $\alpha>2$, purple contours outline cores with $1<\alpha<2$ and orange contours outline cores with $\alpha<1.$
Figure \ref{fig:mom0cores} shows that of the \edit1{43 sub-virial cores (cores with $\alpha<2$), 30 of them are located along the main filaments. 
Of the 22 cores with $\alpha<1$, 20 are located along the main filaments}. 
Cores along the filament are therefore more likely to be gravitationally unstable, as determined by their low virial parameters, than those not along the filaments.  
Observations have shown that bound cores are most often located along filaments that are themselves supercritical \citep{andre_2010,konyves_2015}. 
Indeed, the southern filament in Serpens South has a line mass greater than the critical value \citep{kirk_2013}, and both the southern and northern filaments show evidence for radial infall \citep{kirk_2013, friesen_2013}, as does the central protostellar cluster \citep{tanaka_2013}. 

\edit1{Figure \ref{fig:mom0cores} also shows that the protostellar cluster contains only cores with $\alpha > 1$.} 
This means that cores in the centre currently do not contain sufficient mass to collapse under gravity and form stars. 
This result is in contrast to that found with lower angular resolution, where the overall clump of molecular gas surrounding the central protocluster appears gravitationally unstable following a similar analysis \citep{tanaka_2013}. 
By resolving the dense gas into its small-scale cores, we show for the first time that the protocluster is impacting the ongoing star formation in the nearby environment.  
We investigate further the properties and gravitational stability of the cores with respect to their distance to the young protostellar cluster in 
Section \ref{section:cluster_effect}. 

\subsection{Influence of the protocluster on the surrounding gas}
\label{section:cluster_effect}

We next examine and discuss the influence that the cluster has on the dense cores, and the dense gas more broadly traced by \amm. We calculate the distance of each core's centre from the protostellar cluster centre, defined in Section \ref{sec:kinematics}. 
Figure \ref{fig:parameters_vs_distance_to_centre} shows the kinetic temperature, line width, mass, virial mass and virial parameter as a function of the distance to the center of the cluster for each core (red points). 
In each plot, error bars shown are the standard deviation of these parameters across each core. 

\subsubsection{Gas temperatures}

\edit1{We have shown that} the bulk gas kinetic temperature increases towards the central cluster, and Figure \ref{fig:parameters_vs_distance_to_centre} shows that the average \tkin\ per core follows the same trend. We fit the data with a power law and find $T_K \propto d^{-0.17}$. 
Comparing with other stellar cluster-forming regions, \citet{longmore_2010} find a relationship between the temperature and distance to the centre of the G8.68--0.37 protocluster with a power law index of -0.35, a factor of $\sim 2$ greater than in Serpens South. 
Toward the G28.34+0.06 cloud complex, which contains a young and massive protostellar cluster, \citet{wang_2012} similarly find a dust temperature-radius power law index of -0.36. 
These power laws are in better agreement with theoretical studies that predict a more steep temperature dependence on radius \citep[e.g.,][]{scoville_1976,garay_1999}; however these analyses focused on high mass star-forming regions. 
One explanation for this difference could be due to the fact that G28.34+0.06 and G8.68-0.37 are more massive regions than Serpens South.
More massive stars radiate more strongly at shorter wavelengths,
which are more easily absorbed by intervening dust,
partially explaining the steeper relationship. 
In addition, the dense gas traced by \amm\ in the cluster centre is highly clumped \edit1{(see Figures  \ref{fig:cont_33} and \ref{fig:outflow})}. 
Due to the clumpy nature of the gas, radiation may be able to travel further along some directions, heating the dense gas at further distances and producing a shallower temperature distribution with radius. 

The protostars in Serpens South are low mass, with lower luminosities than in the high mass clusters described above. Alternatively, the heating may be primarily mechanical, due to the interaction of the jets and outflows from the protostars with the surrounding dense gas. Such a mechanism has been proposed for the warmer gas temperatures observed toward the massive protocluster embedded within G28.34+0.06 \citep{wang_2012}. Both highly collimated jets and broad outflows have been identified in Serpens South, extending to several $\times 0.1$~pc from the protocluster centre \citep{Nakamura_2011,teixeira_2012,plunkett_2015_outflows}. Mechanical heating would then also be spread over the extent where outflows are interacting with the dense gas. 

\begin{figure}
    \includegraphics[width=0.48\textwidth]{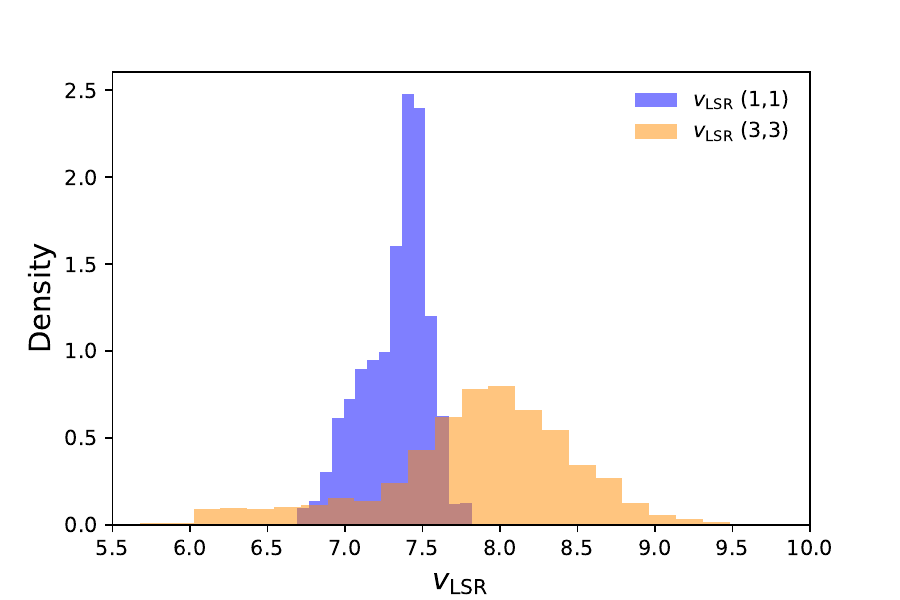}
    \includegraphics[width=0.48\textwidth]{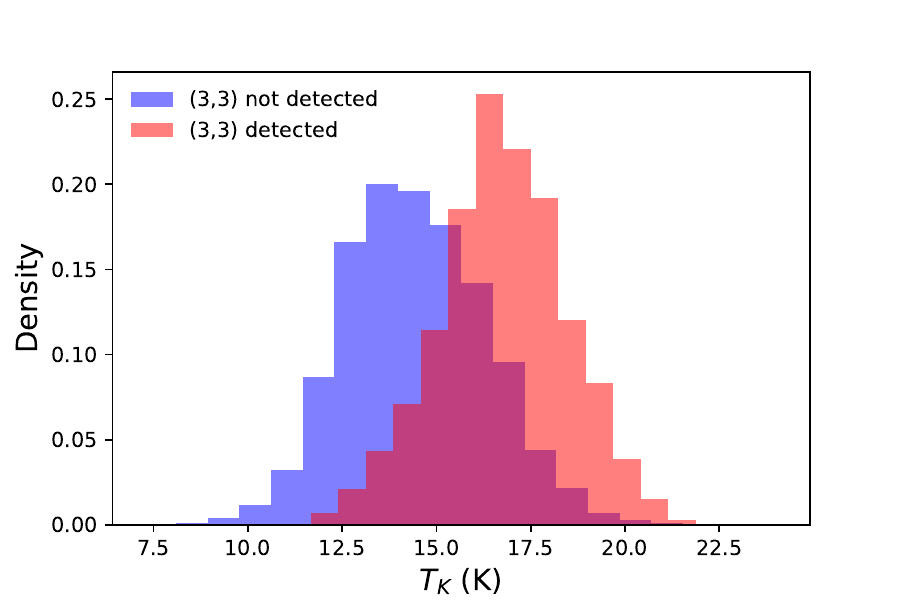}
    \caption{Top: Density distribution of \amm\ (1,1) and (3,3) \vlsr\ (blue and orange histograms, respectively) where both species are detected within the box highlighted in Figure \ref{fig:cont_33}. The \amm\ (3,3) \vlsr\ is significantly offset from the \amm\ (1,1) \vlsr, highlighting its non-thermal excitation in this region. Bottom: A comparison of the distribution of gas temperatures \tkin\ where \amm\ (3,3) is detected (red) and not detected (blue) near the Serpens South protocluster, within the box highlighted in Figure \ref{fig:cont_33}. \label{fig:tk_33_compare}}
\end{figure}

\edit1{We show in Figure \ref{fig:outflow}d) the \amm\ (3,3) integrated intensity contours over a map of the collimated $^{12}$CO outflow driven by the Class 0/I protostellar source CARMA-7 \citep{plunkett_2015_epi}. The $^{12}$CO data were observed with ALMA (resolution 0\farcs9), and the published map extent is narrower than our \amm\ observations. Nonetheless, the brightest \amm\ (3,3) emission is coincident with the collimated outflow, and peaks along the outflow axis to the north where \namm\ actually declines.}
\edit1{The \amm\ (3,3) emission} is likely non-thermally excited given its lower state energy $E_L = 85$~K \citep{pickett_1998} and the measured gas temperatures \tkin\ $\lesssim 20$~K. Indeed, excitation and masing of the \amm\ (3,3) line has been reported previously along molecular outflows and toward high mass star-forming regions \citep{zhang_1995,zhang_1999,tursun_2022}. In support of this argument, the line \vlsr\ and \sigv\ between the (1,1) and (3,3) transitions are significantly different. We show in Figure \ref{fig:tk_33_compare} (top) the distribution of \vlsr\ measured in both transitions, in pixels where both lines are detected well. The \amm\ (3,3) \vlsr\ extends to both lower and higher values compared with \amm\ (1,1), as expected if the \amm\ (3,3) is tracing interaction regions between outflows and dense gas, where some dense gas may become entrained in the flow. As a final check, we use typical values of \namm\ and \tkin\ from the \amm\ (1,1) and (2,2) fits, and the \amm\ (3,3) Gaussian line widths, to estimate an \amm\ (3,3) line amplitude if the emission were thermally excited at a density $n \sim 10^6$~cm$^{-3}$ via a radiative transfer calculation \citep[RADEX;][]{vandertak_2007}. The calculation predicts an \amm\ (3,3) line amplitude of $\lesssim 0.2$~K. In contrast, we detect \amm\ (3,3) with line amplitudes $\sim 0.4 - 0.8$~K, much greater than expected via radiative transfer analysis. 

We compare directly the gas temperatures in pixels where \amm\ (3,3) is detected with a SNR $> 3$ with those where \amm\ (3,3) is not detected in Figure \ref{fig:tk_33_compare} (bottom). We focus only on the area surrounding the protocluster identified by the box in Figure \ref{fig:cont_33}. We find that within the cluster itself, the gas temperature distribution where \amm\ (3,3) is detected differs significantly, and is warmer on average, than where \amm\ (3,3) is not detected. This shows that the warmest gas temperatures in Serpens South correlate with the locations where jets and outflows impact the dense gas, indicative of mechanical heating. We note, however, that even outside of the \amm\ (3,3) detections, the gas temperatures near the cluster remain higher than in the filament by $\sim 2$~K, and radiative heating is likely also playing a role. 

\subsubsection{Velocity dispersions}

We also see a trend with increasing mean core $\sigma_v$ towards the central cluster, shown in the second panel of Figure \ref{fig:parameters_vs_distance_to_centre}. 
To highlight the trend more clearly, we bin the data in bins of width $\sim 0.1$~pc in distance to the central cluster (black points). 
The uncertainty is each of these binned points is taken to be the standard deviation of values within each bin. We see that there is spread in the data for cores within 0.4~pc of the central cluster, but there is significantly less spread for cores further than 0.4~pc. Furthermore, we see that the average values in each bin are larger within $\sim 0.4$~pc of the cluster center. The green line in this panel shows the line width that we would expect if the non-thermal component of the line width were constant and equal to $\sim 0.15$ (the mean value for cores further than 0.5~pc from the central cluster), and we calculate the thermal component following the trend in \tkin\ in the top panel. This shows that most of the increase in line width towards the central cluster is in fact due to the non-thermal component of the line width, rather than the increase in temperature towards the central cluster.  
At larger distances ($\gtrsim 0.5$~pc) from the cluster, the measured line widths are consistent with a constant non-thermal component. 

In Serpens South, increased velocity dispersion in the dense gas may be driven by several factors, such as the interaction of the protostellar jets and outflows with the dense gas, or via accretion onto the cluster via the dense filaments. \citet{kirk_2013} measured a velocity gradient along the southern filament toward the protocluster, with a corresponding mass accretion rate $\dot{M} \sim 50$~M$_\odot$~Myr$^{-1}$ (corrected for $d = 429$~pc). \citet{matzner_2015} show that accretion-driven turbulence can produce observed line widths following $\sigma_{acc}^3 \sim G\dot{M}$, giving $\sigma_{acc} \sim 0.6$~km~s$^{-1}$ for the measured mass accretion rate, which is similar to the \amm\ velocity dispersions we observe in the central cluster. 
\edit1{\citet{trevino-morales_2019} find that flow from filaments onto a central cluster in the Monoceros R2 star forming region results in larger line widths in the central hub. }
In Figure \ref{fig:vlsr_kde}, the \signt\ values may show more of a jump in magnitude near the protocluster, as compared with the smooth increase in \tkin\ with decreasing distance from the cluster centre, but with significant scatter. 

We argue, however, that the increased gas temperatures near the cluster are primarily due to mechanical heating via jets and outflows, and Figure \ref{fig:psd} shows a strong correlation between \tkin\ and \signt. This suggests that the same mechanism is responsible for both the increased \tkin\ and \signt\ values. In numerical models, \citet{nakamura_2007} show that collimated jets and outflows from young protoclusters can indeed drive significant turbulent motions in the dense gas, leading to clumps that remain closer to dynamical equilibrium. \edit1{Collimated outflows in particular are more efficient in driving turbulence. Figure \ref{fig:outflow} shows directly the impact of the collimated outflow driven by CARMA-7 on \signt\ and \tkin\ in the Serpens South protocluster.}
\edit1{Serpens South additionally contains wider-angle outflows with extents of $\sim 0.3$~pc relative to the protocluster center \citep{Nakamura_2011}, broadly matching the region where we find increased \signt\ and \tkin. }

It is likely that both \edit1{accretion and mechanical effects} are playing a role \edit1{in increasing the velocity dispersion in the dense gas}, perhaps at different locations within the cluster. \edit1{A transition from trans-sonic to supersonic \signt\ is seen, for example, where the southern filament connects to the dense gas in the cluster (Figure \ref{fig:outflow}), but this overlaps with the extents of known outflows.} A detailed examination of where the filamentary flows impact the cluster gas, compared with high resolution outflow maps, may better disentangle these effects, but is beyond the scope of this paper. The end result, however, is that both the extended dense gas near the cluster centre, as well as the dense gas within the compact cores, show much greater \signt\ values than in the more distant filaments. 

\subsubsection{Effects on core stability}

The lower three panels of Figure \ref{fig:parameters_vs_distance_to_centre} show the measured core masses, core virial masses, and core virial parameters. We similarly bin the data in these panels as for the velocity dispersion with the distance to the central cluster. 

The third panel of Figure \ref{fig:parameters_vs_distance_to_centre} shows that there is no clear trend in the core mass as a function of distance to the central cluster.
In the fourth panel, we see that on average, core virial masses are greater within $\sim 0.4$~pc of the cluster centre, compared with those further than 0.4pc from the centre. Within $\sim 0.4$~pc of the cluster centre cores have a mean virial mass of 2.7~\msun\ \edit1{with a standard deviation of 1.9~\msun} whereas cores further than  $\sim 0.4$~pc of the cluster centre have a mean virial mass of 1.5~\msun \edit1{with a standard deviation of only 0.7~\msun}. 
\edit1{While there are several cores with virial masses significantly above the typical value, excluding these points does not significantly affect the above results.}
From the first two panels in Figure \ref{fig:parameters_vs_distance_to_centre}, we see that the cluster impacts on the dense gas are in both heating and an increase in non-thermal motions. 
These effects combine to produce the higher virial masses that we see towards the cluster centre, however it is the non-thermal velocity dispersion that dominates the increase in virial mass. The mean non-thermal component of the velocity dispersion is $\sim1.5$ times greater for cores within 0.4~pc of the cluster centre than for cores further than 0.4~pc. The mean temperature of the cores within 0.4~pc is only $\sim1.2$ times greater than the mean temperature of cores further from the centre. As a result, the increase in temperatures only causes the virial masses to increase by 5\% whereas the increases in the non-thermal component of the velocity dispersion increase the virial masses by 55\%. This indicates that the increase in virial mass is primarily due to the non-thermal component of the velocity dispersion.

As a result, we see in the bottom panel of Figure \ref{fig:parameters_vs_distance_to_centre} that, on average, the core virial parameters also increase towards the central cluster. 
This follows from the observed increase in core kinetic temperature and velocity dispersion toward the central cluster, and hence the increase in the virial mass, 
since there is no corresponding trend with core mass. 
This trend indicates that cores near the protocluster tend to be more stable and can be supported with higher masses before becoming gravitationally unstable.

Within $\sim 0.4$~pc of the cluster centre, cores have a mean virial parameter of \edit1{3.3 with a standard deviation of 2.7} whereas for cores further than 0.4~pc from the central cluster have a mean virial parameter of \edit1{1.9 with a standard deviation of 1.7}. 
\edit1{Cores in the protocluster are therefore more likely to be super-virial, with virial parameters $\alpha$ increased by a factor $\sim 1.7$ near the cluster centre.}
If the gas-to-stellar mass efficiency in the cores is constant, this means that the next generation of stars in the cluster will be higher mass than those further from the central cluster. 
In addition, because many cores in the central cluster have $\alpha>2$, they will need to accrete more mass in order to become gravitationally unstable and eventually form stars. 

\subsection{Future of star formation in Serpens South}

As described previously, Serpens South is a forming protostellar cluster, with a relatively large fraction of embedded sources indicating its young age. Our results thus highlight the impact of the first generation of stars within a young cluster on the surrounding dense gas that has yet to form stars. We find strong trends of increasing gas temperatures and non-thermal line widths with decreasing distance to the cluster centre. Both parameters correlate positively with the surface density of YSOs, and strongly correlate with each other, which we argue is a signature of mechanical heating and energy injection via protostellar outflows driven within the protocluster. 

Cores within the \edit1{gravitationally supercritical (M$_{vir}$ $<$ M$_{core}$)} filaments are more likely to themselves appear gravitationally unstable following a virial analysis, as seen toward other star-forming clouds. Near the cluster centre, however, the core masses are similar on average, but the increased gas temperatures and non-thermal line widths result in high virial masses and thus virial parameters. The increased virial mass is driven primarily by the increased velocity dispersion rather than the temperature. Toward Serpens South, gas temperatures increase by only $\sim 5-7$~K near the protocluster. In higher mass star-forming regions, gas temperatures may increase more significantly and have greater impact on the potential stability of embedded cores. Some studies have found slightly greater dense gas temperatures, measured via \amm\ emission, near more massive, young clusters \citep{wang_2012,fontani_2012,tursun_2020,kohno_2022}. Other studies of infrared dark clouds, potential sites of future cluster formation, found little difference in the gas temperatures toward star-forming and starless regions \citep{ragan_2011}.  

Toward Serpens South, core virial masses increase by a factor of $\sim 2$ between cores in the filaments and cores near the central protocluster. In order to collapse, these cores must either accrete more mass from their environment, or their turbulence must decay. If cores become unstable via mass accretion, this next generation of stars will be of higher mass, on average, than the current generation of protostars, assuming a similar fraction of core mass ends up in the nascent star.   

Can the cores accrete additional mass? The central region itself contains $\sim 330$~$M_\odot$ of gas, of which $\sim 27$~\% is in cores. There is therefore a central reservoir within which the cores are embedded, from which they may accrete material. Furthermore, if we interpret the velocity gradients along the filaments as inward flow toward the cluster, there is an additional source of mass accretion. In this interpretation, the complex must be inclined at some angle $i$ with respect to the plane of the sky. As in \citet{kirk_2013}, we assume an angle in the range $30^{\circ} < i < 60^{\circ}$ is a reasonable range for the inclination. Given the measured velocity gradient of $0.7 \pm 0.1$~km~s$^{-1}$~pc$^{-1}$, with this range of inclinations the gas at $\sim 0.5$~pc from the cluster centre would take between $\sim 0.7$~Myr to \edit1{$\sim 1.4$~Myr} to reach and accrete onto the cluster. This is the furthest extent along the southern filament where we see the smooth velocity gradient in Figure \ref{fig:vlsr_kde}. At greater distances, it is not clear that the gas will flow toward the cluster, and instead is likely more closely bound to the local clumps within the filament. To the north, the \edit1{large-scale} velocity gradient breaks at a distance $\sim 0.3$~pc, with a consequently shorter accretion time. This limits the mass reservoir and total timescale available for star formation in the central cluster. Furthermore, while protostellar outflows are not yet sufficient to disrupt the larger star-forming clump \citep{Nakamura_2011,plunkett_2015_epi}, as more stars form the feedback effects should eventually be sufficient to halt accretion \citep{matzner_2015}. 

Figure \ref{fig:vlsr_kde} further shows that cores are largely moving with the surrounding dense gas. Multiple cores within the filaments will thus follow the gas flow toward the cluster. Cores within the filaments are already more likely to be gravitationally bound (Section \ref{section:stability_discussion}). Given a median mass $M \sim 0.8$~M$_\odot$ and median radius $R \sim 0.01$~pc, the free-fall time for a typical core is a few times $10^4$ to $\sim 10^5$~years. These bound, filamentary cores will thus likely form stars before the cores accrete onto the central cluster. The central cluster will therefore accrete both young protostars and additional molecular gas over the next $\sim 1-2$~Myr, increasing the number of cluster members, as well as providing an additional mass reservoir for the current cluster cores to gain mass and potentially become gravitationally unstable and collapse to form protostars themselves. 

\section{Summary}\label{section:summary}
We have used 5\arcsec\ angular resolution data from the Very Large Array to map \amm\ (1,1), (2,2), and (3,3) emission toward the nearby, young, clustered star forming region, Serpens South.
\begin{enumerate}

    \item We perform hyperfine line fitting of the \amm\ (1,1) and (2,2) emission to obtain maps of the gas temperature, \amm\ column density, velocity dispersion and line of sight velocity towards Serpens South.
    
    \item Using a dendrogram analysis, we examine the hierarchical structure of the dense gas in Serpens South and we classify dense gas cores as the upper-level structures obtained from this analysis. We locate 108 dense cores in Serpens South.
    
    \item We assess the stability of the 94 dense cores using a virial analysis. We find that most cores \edit1{(54\%)} are super-virial with $\alpha>2$. Most gravitationally bound cores are located along the filaments. Core masses near the cluster centre are similar on average to those in the filaments but have greater virial parameters, leading to $\alpha > 2$ toward most cores within $\sim 0.4$~pc of the YSO cluster centre.
    
    \item We find that dense gas temperatures, including mean temperatures within the cores, increase significantly with decreasing distance to the protostellar cluster centre, but the dependence of temperature on radius is more shallow than found toward higher mass star forming regions. 
    
    \item We find a general trend of non-thermal line width increasing towards the central cluster. We argue that the increase in non-thermal motions, and the increase in gas temperatures, are both driven primarily by the impact of known protostellar jets and outflows on the dense gas near the protocluster. 
    
    \item The increased velocity dispersions within the cores drive an increase in the measured virial parameter towards the cluster centre. Within 0.4~pc of the centre, cores have a mean virial parameter \edit1{$\sim$ 1.7} times greater than those further than 0.4~pc. All cores within 0.4~pc of the cluster are super-virial and will be able to form stars (with consequently higher masses) only if they accrete more mass, or their non-thermal motions dissipate. 
    
    \item Bound cores embedded in the filaments will likely form stars before accreting onto the central cluster. This will increase the number of cluster members within $\sim$1--2Myr.
    
\end{enumerate}

\begin{acknowledgments}

The authors thank the anonymous referee for thoughtful comments that improved the paper. 
The authors thank C. Matzner and A. Sills for fruitful discussions and feedback, and A. Plunkett for providing the $^{12}$CO ALMA data for the CARMA-7 outflow. 
The University of Toronto operates on the traditional land of the Huron-Wendat, the Seneca, and most recently, the Mississaugas of the Credit River; we are grateful to have the opportunity to work on this land. 
The National Radio Astronomy Observatory is a facility of the National Science Foundation operated under cooperative agreement by Associated Universities, Inc.
The Green Bank Observatory is a facility of the National Science Foundation operated under cooperative agreement by Associated Universities, Inc.

\edit1{\amm\ data cubes, integrated intensity maps, and line fit parameter maps are publicly available at the CADC via the following link: \dataset[10.11570/24.0006]{https://www.canfar.net/citation/landing?doi=24.0006}}

\end{acknowledgments}

%

\vspace{5mm}
\facilities{VLA, GBT}


\software{astropy \citep{astropy:2013,astropy:2018},  
          astrodendro \citep{rosolowsky_2008},
          scipy \citep{2020SciPy-NMeth},
          pyspeckit \citep{pyspeckit_2011,pyspeckit_2022}
          }



\appendix
\section{Core properties}


\startlongtable

\begin{deluxetable*}{cccccccccc} 
\tabletypesize{\footnotesize}
\tablecaption{Core properties. \label{tab:core_properties}}
\tablecolumns{10} 
\tablewidth{0pt}
\tablehead{
\colhead{ID} & \colhead{RA DEC} & \colhead{R} & \colhead{$N$(NH$_3$)} & \colhead{$T_\mathrm{K}$} & \colhead{$\sigma_v$} & \colhead{$v_\mathrm{LSR}$} & \colhead{$M_\mathrm{core}$} & \colhead{$M_\mathrm{vir}$} & \colhead{$\alpha_\mathrm{vir}$} \\
 &\colhead{($\mathrm{J2000})$} & \colhead{$(\mathrm{pc})$} & \colhead{$\times 10 ^ {14} (\textrm{cm}^{-2})$} &\colhead{$\mathrm{(K)}$} & \colhead{(km s$^{-1}$)} & \colhead{(km s$^{-1}$)} & \colhead{(M$_{\odot}$)} & \colhead{(M$_{\odot}$)} & } 

\startdata
1 & 18:29:50.40 -01:57:22.71 & 0.016 & 2.1 & 10.1(1.3) & 0.25(0.04) & 7.55(0.06) & 1.81(0.13) & 1.7 & 0.9(0.1) \\
2 & 18:29:51.35 -01:57:50.73 & 0.015 & 1.9 & 10.4(1.1) & 0.22(0.02) & 7.57(0.05) & 1.85(0.1) & 1.5 & 0.8(0.0) \\
3 & 18:29:51.38 -01:57:29.97 & 0.014 & 2.3 & 10.4(1.4) & 0.25(0.03) & 7.68(0.05) & 1.85(0.12) & 1.6 & 0.9(0.1) \\
4 & 18:29:51.56 -01:56:41.47 & 0.009 & - & - & - & - & - & - & - \\
5 & 18:29:52.18 -01:57:57.78 & 0.011 & 2.3 & 9.9(1.1) & 0.23(0.04) & 7.65(0.07) & 1.23(0.16) & 1.1 & 0.9(0.1) \\
6 & 18:29:52.39 -01:56:25.86 & 0.008 & 1.1 & 13.0(0.7) & 0.14(0.01) & 6.9(0.02) & 0.28(0.02) & 0.6 & 2.1(0.1) \\
7 & 18:29:52.56 -01:58:10.69 & 0.017 & 2.3 & 10.8(1.8) & 0.23(0.04) & 7.52(0.05) & 3.15(0.22) & 1.8 & 0.6(0.0) \\
8 & 18:29:54.68 -01:56:27.25 & 0.008 & 1.7 & 10.2(0.9) & 0.31(0.04) & 7.08(0.05) & 0.26(0.03) & 1.2 & 4.6(0.5) \\
9 & 18:29:54.73 -01:59:16.73 & 0.007 & 2.4 & 11.6(1.2) & 0.33(0.09) & 7.43(0.03) & 0.4(0.01) & 1.2 & 2.9(0.1) \\
10 & 18:29:54.87 -02:00:20.18 & 0.006 & - & - & 0.45(0.06) & 7.38(0.07) & - & - & - \\
11 & 18:29:55.18 -01:56:25.97 & 0.007 & 1.7 & 10.2(0.9) & 0.23(0.04) & 7.25(0.06) & 0.21(0.02) & 0.7 & 3.4(0.3) \\
12 & 18:29:55.88 -02:01:07.15 & 0.016 & 3.9 & 11.9(0.8) & 0.27(0.03) & 7.88(0.04) & 3.13(0.22) & 2.1 & 0.7(0.0) \\
13 & 18:29:56.02 -01:58:26.14 & 0.012 & 4.9 & 10.3(0.4) & 0.16(0.03) & 7.47(0.03) & 1.74(0.14) & 0.9 & 0.5(0.0) \\
14 & 18:29:56.25 -02:00:20.43 & 0.021 & 1.9 & 11.9(1.7) & 0.4(0.08) & 7.31(0.07) & 3.26(0.42) & 5.0 & 1.5(0.2) \\
15 & 18:29:56.29 -01:56:10.68 & 0.013 & 2.0 & 9.6(1.4) & 0.18(0.02) & 7.33(0.02) & 0.88(0.1) & 0.9 & 1.1(0.1) \\
16 & 18:29:56.72 -01:59:58.45 & 0.007 & - & - & 0.46(0.04) & 7.27(0.04) & - & - & - \\
17 & 18:29:56.72 -02:00:40.42 & 0.008 & 2.5 & 12.2(1.3) & 0.49(0.03) & 7.4(0.02) & 0.39(0.02) & 2.5 & 6.5(0.4) \\
18 & 18:29:57.00 -02:00:30.39 & 0.013 & 1.9 & 14.0(2.1) & 0.42(0.07) & 7.38(0.04) & 0.86(0.07) & 3.3 & 3.9(0.3) \\
19 & 18:29:57.06 -01:56:50.18 & 0.011 & 4.2 & 9.9(0.6) & 0.2(0.02) & 7.47(0.02) & 0.28(0.02) & 0.9 & 3.3(0.2) \\
20 & 18:29:57.06 -02:01:09.00 & 0.011 & 3.5 & 12.3(1.1) & 0.28(0.08) & 7.88(0.06) & 1.48(0.11) & 1.5 & 1.0(0.1) \\
21 & 18:29:57.42 -01:58:27.87 & 0.008 & 5.8 & 10.1(0.7) & 0.21(0.01) & 7.57(0.01) & 0.38(0.02) & 0.7 & 1.9(0.1) \\
22 & 18:29:57.43 -01:58:10.90 & 0.016 & 6.6 & 10.1(0.7) & 0.2(0.02) & 7.52(0.02) & 4.39(0.32) & 1.4 & 0.3(0.0) \\
23 & 18:29:57.46 -01:59:29.95 & 0.013 & 2.5 & 10.9(1.0) & 0.24(0.05) & 7.47(0.07) & 2.55(0.17) & 1.4 & 0.6(0.0) \\
24 & 18:29:57.56 -02:01:17.27 & 0.008 & 3.3 & 11.8(0.8) & 0.23(0.03) & 7.87(0.03) & 0.77(0.03) & 0.8 & 1.1(0.0) \\
25 & 18:29:57.58 -02:00:44.53 & 0.007 & 2.1 & 9.4(0.8) & 0.44(0.08) & 7.53(0.05) & 0.26(0.04) & 1.8 & 7.0(1.0) \\
26 & 18:29:57.68 -02:00:01.83 & 0.01 & 1.6 & 11.5(1.2) & 0.35(0.07) & 7.21(0.05) & 0.79(0.06) & 1.8 & 2.3(0.2) \\
27 & 18:29:57.72 -01:57:12.02 & 0.024 & 6.3 & 9.7(0.7) & 0.22(0.02) & 7.54(0.04) & 8.61(0.77) & 2.3 & 0.3(0.0) \\
28 & 18:29:57.78 -01:57:33.40 & 0.013 & 6.9 & 10.4(0.9) & 0.22(0.01) & 7.62(0.02) & 3.15(0.28) & 1.2 & 0.4(0.0) \\
29 & 18:29:58.04 -01:59:48.08 & 0.008 & 1.7 & 11.7(1.2) & 0.27(0.05) & 7.35(0.04) & 0.75(0.05) & 1.1 & 1.4(0.1) \\
30 & 18:29:58.45 -02:01:13.70 & 0.02 & 4.0 & 11.8(1.3) & 0.55(0.05) & 7.52(0.13) & 6.77(0.66) & 8.0 & 1.2(0.1) \\
31 & 18:29:58.89 -02:00:57.57 & 0.008 & 2.9 & 12.6(0.4) & 0.26(0.03) & 7.75(0.03) & 0.6(0.03) & 1.1 & 1.8(0.1) \\
32 & 18:29:59.27 -01:59:12.19 & 0.006 & 1.7 & 10.0(1.5) & 0.26(0.02) & 7.31(0.05) & 0.3(0.03) & 0.7 & 2.4(0.3) \\
33 & 18:29:59.32 -02:02:37.15 & 0.01 & 3.1 & 15.3(0.4) & 0.45(0.08) & 7.4(0.11) & 1.2(0.11) & 3.0 & 2.5(0.2) \\
34 & 18:29:59.66 -02:02:50.44 & 0.011 & 2.1 & 12.7(1.2) & 0.42(0.1) & 7.29(0.12) & 0.91(0.06) & 3.0 & 3.3(0.2) \\
35 & 18:29:59.74 -02:00:52.54 & 0.011 & 3.0 & 11.3(1.3) & 0.35(0.04) & 7.72(0.04) & 1.4(0.07) & 2.1 & 1.5(0.1) \\
36 & 18:29:59.91 -02:03:05.48 & 0.01 & 1.6 & 14.5(2.0) & 0.32(0.04) & 7.22(0.04) & 0.52(0.04) & 1.7 & 3.2(0.2) \\
37 & 18:30:00.10 -02:01:25.34 & 0.01 & 1.6 & 12.8(1.8) & 0.37(0.04) & 7.59(0.05) & 0.56(0.03) & 2.0 & 3.6(0.2) \\
38 & 18:30:00.10 -02:03:23.26 & 0.008 & 2.3 & 14.9(1.0) & 0.36(0.03) & 7.26(0.03) & 0.51(0.07) & 1.5 & 3.0(0.4) \\
39 & 18:30:00.22 -02:01:58.36 & 0.008 & 1.8 & 14.3(1.6) & 0.26(0.03) & 7.01(0.04) & 0.4(0.03) & 1.0 & 2.6(0.2) \\
40 & 18:30:00.41 -01:58:18.69 & 0.008 & 2.0 & 9.7(0.8) & 0.31(0.03) & 7.73(0.06) & 0.56(0.04) & 1.3 & 2.3(0.1) \\
41 & 18:30:00.41 -01:59:58.81 & 0.033 & 1.8 & 11.4(1.4) & 0.19(0.07) & 7.29(0.09) & 8.72(1.11) & 2.9 & 0.3(0.0) \\
42 & 18:30:01.15 -02:01:32.71 & 0.012 & 3.1 & 12.1(1.3) & 0.56(0.04) & 7.58(0.06) & 1.69(0.13) & 4.8 & 2.9(0.2) \\
43 & 18:30:01.35 -02:04:42.75 & 0.008 & - & - & - & - & - & - & - \\
44 & 18:30:01.37 -02:05:03.37 & 0.006 & - & - & - & - & - & - & - \\
45 & 18:30:01.54 -02:02:20.67 & 0.006 & - & - & 0.45(0.02) & 7.37(0.06) & - & - & - \\
46 & 18:30:01.79 -02:02:03.54 & 0.009 & 3.3 & 16.5(0.5) & 0.38(0.05) & 7.58(0.03) & 0.83(0.04) & 2.0 & 2.4(0.1) \\
47 & 18:30:01.80 -02:01:42.72 & 0.008 & 2.9 & 13.0(1.4) & 0.5(0.05) & 7.53(0.15) & 0.63(0.02) & 2.6 & 4.1(0.2) \\
48 & 18:30:01.81 -02:00:48.03 & 0.006 & 2.5 & 13.6(1.0) & 0.34(0.03) & 7.12(0.05) & 0.27(0.01) & 1.1 & 4.1(0.1) \\
49 & 18:30:01.88 -02:03:56.67 & 0.021 & 2.7 & 13.7(0.8) & 0.33(0.04) & 7.02(0.09) & 4.12(0.22) & 3.9 & 0.9(0.1) \\
50 & 18:30:01.90 -02:01:33.54 & 0.008 & - & - & 0.53(0.07) & 7.62(0.08) & - & - & - \\
51 & 18:30:01.98 -02:01:21.03 & 0.01 & 3.2 & 11.8(1.0) & 0.48(0.03) & 7.74(0.05) & 1.01(0.04) & 3.1 & 3.0(0.1) \\
52 & 18:30:02.26 -02:02:49.77 & 0.009 & 5.1 & 16.4(0.5) & 0.46(0.02) & 7.33(0.07) & 0.54(0.01) & 2.6 & 4.8(0.1) \\
53 & 18:30:02.32 -02:02:40.79 & 0.007 & 5.3 & 15.1(0.4) & 0.4(0.04) & 6.95(0.03) & 0.41(0.02) & 1.8 & 4.4(0.2) \\
54 & 18:30:02.45 -02:05:26.55 & 0.013 & 1.6 & 11.7(1.1) & 0.27(0.04) & 7.68(0.06) & 0.82(0.08) & 1.6 & 2.0(0.2) \\
55 & 18:30:02.46 -02:02:27.11 & 0.008 & 5.0 & 16.7(1.5) & 0.48(0.05) & 7.36(0.18) & 0.41(0.02) & 2.7 & 6.7(0.3) \\
56 & 18:30:02.77 -02:00:26.28 & 0.011 & 2.2 & 12.0(0.8) & 0.39(0.07) & 7.2(0.06) & 0.77(0.06) & 2.6 & 3.4(0.3) \\
57 & 18:30:02.98 -02:01:00.37 & 0.032 & 3.6 & 12.8(1.2) & 0.44(0.08) & 7.37(0.17) & 12.3(0.98) & 9.0 & 0.7(0.1) \\
58 & 18:30:03.03 -02:03:22.16 & 0.017 & 5.4 & 13.3(0.6) & 0.41(0.07) & 7.07(0.09) & 5.24(0.27) & 4.2 & 0.8(0.0) \\
59 & 18:30:03.14 -02:01:31.95 & 0.008 & 1.7 & 15.1(2.1) & 0.41(0.03) & 7.44(0.04) & 0.41(0.03) & 2.0 & 4.8(0.4) \\
60 & 18:30:03.14 -02:04:37.35 & 0.008 & 3.2 & 12.7(1.1) & 0.28(0.02) & 6.73(0.02) & 0.56(0.03) & 1.0 & 1.9(0.1) \\
61 & 18:30:03.36 -02:04:27.27 & 0.007 & 2.9 & 12.6(0.3) & 0.21(0.03) & 6.69(0.02) & 0.3(0.01) & 0.7 & 2.3(0.0) \\
62 & 18:30:03.37 -02:02:37.59 & 0.008 & - & - & 0.42(0.06) & 7.46(0.05) & - & - & - \\
63 & 18:30:03.40 -01:59:38.41 & 0.02 & 1.6 & 10.9(1.7) & 0.24(0.07) & 7.26(0.12) & 2.81(0.32) & 2.3 & 0.8(0.1) \\
64 & 18:30:03.56 -02:05:52.30 & 0.007 & - & - & 0.29(0.09) & 7.46(0.04) & - & - & - \\
65 & 18:30:03.63 -02:02:06.01 & 0.013 & 3.2 & 16.4(0.9) & 0.44(0.09) & 7.73(0.06) & 2.17(0.14) & 3.8 & 1.7(0.1) \\
66 & 18:30:03.63 -02:05:28.24 & 0.008 & 1.7 & 12.6(1.4) & 0.45(0.04) & 7.35(0.06) & 0.34(0.02) & 2.4 & 7.2(0.5) \\
67 & 18:30:03.82 -02:03:52.38 & 0.012 & 3.3 & 13.2(0.5) & 0.37(0.11) & 6.83(0.08) & 1.1(0.04) & 2.7 & 2.5(0.1) \\
68 & 18:30:04.06 -02:05:23.55 & 0.01 & 1.7 & 11.0(1.1) & 0.47(0.06) & 7.17(0.08) & 0.4(0.03) & 3.1 & 7.8(0.7) \\
69 & 18:30:04.11 -02:03:10.87 & 0.021 & 4.6 & 17.1(1.5) & 0.63(0.1) & 7.44(0.18) & 10.51(0.72) & 11.3 & 1.1(0.1) \\
70 & 18:30:04.30 -02:00:18.34 & 0.011 & - & - & - & - & - & - & - \\
71 & 18:30:04.47 -01:59:57.46 & 0.013 & 1.4 & 11.4(1.4) & 0.16(0.06) & 7.16(0.08) & 1.19(0.12) & 1.0 & 0.8(0.1) \\
72 & 18:30:04.59 -02:05:44.34 & 0.014 & 1.5 & 11.3(1.3) & 0.41(0.04) & 7.3(0.06) & 0.86(0.09) & 3.4 & 4.0(0.4) \\
73 & 18:30:04.82 -02:05:00.90 & 0.008 & 2.0 & 12.1(0.6) & 0.23(0.03) & 6.88(0.02) & 0.38(0.01) & 0.9 & 2.3(0.1) \\
74 & 18:30:05.07 -02:01:21.95 & 0.017 & 2.4 & 14.9(0.0) & 0.45(0.11) & 7.34(0.11) & 1.73(0.0) & 5.0 & 2.9(0.0) \\
75 & 18:30:05.19 -02:00:45.03 & 0.006 & 2.5 & 10.9(1.2) & 0.64(0.04) & 7.64(0.08) & 0.21(0.01) & 3.3 & 15.4(0.8) \\
76 & 18:30:05.25 -02:04:44.66 & 0.007 & 2.3 & 12.7(1.7) & 0.51(0.08) & 6.87(0.04) & 0.36(0.04) & 2.7 & 7.3(0.7) \\
77 & 18:30:05.42 -02:05:56.81 & 0.011 & 1.5 & 12.2(1.4) & 0.47(0.08) & 7.41(0.05) & 0.45(0.04) & 3.3 & 7.2(0.7) \\
78 & 18:30:05.56 -02:03:09.48 & 0.008 & 3.1 & 13.4(0.7) & 0.49(0.04) & 7.28(0.1) & 0.92(0.04) & 2.8 & 3.0(0.1) \\
79 & 18:30:05.73 -02:01:53.88 & 0.014 & 1.6 & 12.4(1.3) & 0.3(0.05) & 7.74(0.05) & 1.01(0.09) & 2.1 & 2.1(0.2) \\
80 & 18:30:05.83 -02:00:43.99 & 0.007 & 2.7 & 11.7(1.5) & 0.65(0.07) & 7.54(0.04) & 0.32(0.04) & 3.9 & 12.2(1.4) \\
81 & 18:30:06.00 -02:00:58.74 & 0.011 & 1.6 & 11.1(1.7) & 0.25(0.16) & 6.91(0.12) & 0.63(0.1) & 1.6 & 2.5(0.4) \\
82 & 18:30:06.20 -02:05:11.02 & 0.02 & 2.5 & 12.2(1.0) & 0.3(0.05) & 6.97(0.07) & 2.71(0.2) & 3.0 & 1.1(0.1) \\
83 & 18:30:06.29 -02:03:51.66 & 0.013 & 1.6 & 12.9(0.9) & 0.23(0.04) & 6.98(0.03) & 1.15(0.09) & 1.5 & 1.3(0.1) \\
84 & 18:30:06.30 -02:01:32.45 & 0.008 & - & - & 0.21(0.09) & 7.58(0.08) & - & - & - \\
85 & 18:30:06.54 -02:05:49.72 & 0.012 & 1.6 & 12.9(1.0) & 0.41(0.07) & 7.16(0.09) & 0.54(0.04) & 3.1 & 5.6(0.4) \\
86 & 18:30:06.58 -02:02:37.83 & 0.006 & - & - & - & - & - & - & - \\
87 & 18:30:06.75 -02:07:04.98 & 0.014 & 1.5 & 11.0(1.5) & 0.19(0.07) & 7.54(0.08) & 0.78(0.1) & 1.2 & 1.6(0.2) \\
88 & 18:30:07.03 -02:05:25.13 & 0.008 & 2.2 & 12.3(0.8) & 0.29(0.02) & 6.98(0.02) & 0.36(0.02) & 1.2 & 3.2(0.1) \\
89 & 18:30:07.27 -02:06:54.10 & 0.009 & 1.3 & 10.7(1.3) & 0.19(0.04) & 7.33(0.04) & 0.31(0.02) & 0.7 & 2.4(0.2) \\
90 & 18:30:07.67 -02:04:37.90 & 0.013 & 2.0 & 12.2(2.1) & 0.39(0.06) & 7.22(0.05) & 0.81(0.08) & 2.8 & 3.5(0.4) \\
91 & 18:30:08.13 -02:06:42.92 & 0.017 & 1.4 & 9.8(1.1) & 0.2(0.03) & 7.1(0.09) & 1.25(0.13) & 1.4 & 1.1(0.1) \\
92 & 18:30:08.15 -02:05:30.14 & 0.014 & 2.3 & 10.5(1.1) & 0.32(0.03) & 6.93(0.04) & 1.1(0.09) & 2.1 & 1.9(0.2) \\
93 & 18:30:08.62 -02:04:46.61 & 0.008 & 1.5 & 10.1(0.4) & 0.38(0.07) & 7.16(0.06) & 0.28(0.01) & 1.8 & 6.2(0.2) \\
94 & 18:30:09.16 -02:04:08.79 & 0.025 & 1.4 & 13.0(1.6) & 0.2(0.03) & 7.04(0.12) & 2.96(0.32) & 2.4 & 0.8(0.1) \\
95 & 18:30:10.68 -02:04:49.24 & 0.013 & - & - & - & - & - & - & - \\
96 & 18:30:10.93 -02:06:03.44 & 0.024 & 4.3 & 10.8(0.5) & 0.23(0.03) & 6.76(0.09) & 4.32(0.33) & 2.4 & 0.6(0.0) \\
97 & 18:30:11.62 -02:03:35.11 & 0.01 & 1.0 & 11.3(0.7) & 0.19(0.03) & 7.37(0.04) & 0.26(0.02) & 0.9 & 3.3(0.2) \\
98 & 18:30:11.64 -02:03:46.88 & 0.014 & - & - & 0.22(0.05) & 7.36(0.07) & - & - & - \\
99 & 18:30:12.02 -02:03:24.94 & 0.008 & 1.1 & 10.6(0.9) & 0.21(0.03) & 7.51(0.09) & 0.16(0.01) & 0.8 & 4.7(0.2) \\
100 & 18:30:12.58 -02:06:51.10 & 0.028 & 3.8 & 11.2(0.5) & 0.24(0.03) & 6.57(0.08) & 6.08(0.29) & 3.1 & 0.5(0.0) \\
101 & 18:30:12.59 -02:06:21.48 & 0.024 & 3.8 & 10.6(0.4) & 0.2(0.02) & 6.61(0.04) & 4.04(0.18) & 2.0 & 0.5(0.0) \\
102 & 18:30:14.41 -02:07:25.93 & 0.02 & 3.1 & 10.9(0.6) & 0.21(0.02) & 6.6(0.04) & 2.83(0.12) & 1.8 & 0.6(0.0) \\
103 & 18:30:15.03 -02:06:55.49 & 0.011 & 3.3 & 10.9(0.5) & 0.25(0.01) & 6.65(0.03) & 0.81(0.02) & 1.3 & 1.6(0.0) \\
104 & 18:30:15.75 -02:06:24.41 & 0.007 & 1.4 & 10.5(1.7) & 0.2(0.03) & 6.58(0.03) & 0.15(0.01) & 0.6 & 4.2(0.4) \\
105 & 18:30:16.13 -02:07:15.51 & 0.011 & 2.7 & 10.3(1.1) & 0.24(0.01) & 6.64(0.04) & 0.31(0.02) & 1.2 & 3.8(0.2) \\
106 & 18:30:16.81 -02:08:33.89 & 0.018 & 1.7 & 11.3(0.9) & 0.24(0.03) & 6.68(0.04) & 0.43(0.03) & 2.0 & 4.5(0.3) \\
107 & 18:30:16.82 -02:07:48.54 & 0.01 & 1.4 & 10.7(0.8) & 0.18(0.01) & 6.64(0.02) & 0.51(0.03) & 0.8 & 1.6(0.1) \\
108 & 18:30:18.95 -02:07:47.95 & 0.023 & 1.3 & 10.6(1.3) & 0.25(0.06) & 6.6(0.05) & 2.67(0.31) & 2.7 & 1.0(0.1) \\
\enddata 

\tablecomments{
All values in the \namm, \tkin, \sigv and \vlsr\ columns are the mean value across the core with bracketed values indicating the 1-$\sigma$ spread in values across each core. Spread in \namm\ values across each core is 0.1 or less. }
\end{deluxetable*}


\section{Continuum sources}

Here we include a table of the detected 1~cm continuum sources as described in Section \ref{sec:continuum}. Continuum source sizes, peak and total fluxes were calculated by fitting 2D Gaussians to the continuum peaks using the \texttt{imfit} task in CASA. Source sizes have not been deconvolved with the VLA beam. 

\begin{rotatetable*}
\movetableright=0.1cm
\begin{deluxetable}{lcccccccc} 
\tabletypesize{\scriptsize} 
\tablecolumns{9} 
\tablewidth{0pt} 
\tablecaption{Gaussian fit results to centimeter continuum sources. \label{tab:continuum} } 
\tablehead{ 
\colhead{Source} & \colhead{Coordinates} & \colhead{Peak} & \colhead{$S_\nu$} & \colhead{$\sigma_\mathrm{maj}$} & \colhead{$\sigma_\mathrm{min}$} & \colhead{P.A.} & 
\colhead{Other names} & \colhead{Classification}\\ 
\colhead{} & \colhead{(J2000)} & \colhead{(mJy/beam)} & \colhead{(mJy)} & \colhead{(\arcsec)} & \colhead{(\arcsec)} & \colhead{(\degr)} & \colhead{} & \colhead{} 
} 
\startdata 
a & 18:30:04.13 -02:03:02.50 & 0.57 (0.04) & 1.10 (0.10) & 07.5 (00.5) & 05.6 (00.3) & 124.7 (08.1) & VLA 12\tablenotemark{a}, MM18\tablenotemark{b}, CARMA-7\tablenotemark{c} & Class 0\tablenotemark{a} \\ 
b & 18:30:03.58 -02:03:09.05 & 0.52 (0.04) & 1.18 (0.11) & 08.2 (00.6) & 06.1 (00.4) & 111.4 (08.4) & VLA 13\tablenotemark{a}, CARMA-6\tablenotemark{c} & Class 0/I\tablenotemark{a} \\ 
c & 18:30:03.35 -02:02:44.65 & 0.12 (0.03) & 0.49 (0.13) & 20.2 (06.1) & 04.5 (00.7) & 120.5 (03.0) & VLA 11\tablenotemark{a}, CARMA-5\tablenotemark{c} & Class I\tablenotemark{a} \\ 
d & 18:30:02.46 -02:02:48.19 & 0.12 (0.03) & 0.25 (0.10) & 08.3 (02.8) & 05.5 (01.4) & 169.3 (23.7) & & \\ 
e & 18:30:01.31 -02:03:42.71 & 0.32 (0.03) & 0.43 (0.07) & 05.9 (00.7) & 04.8 (00.5) & 119.1 (18.8) & VLA 17\tablenotemark{a} & Flat/Class II\tablenotemark{a} \\ 
f & 18:30:05.87 -02:01:44.57 & 0.14 (0.03) & 0.31 (0.10) & 11.4 (03.4) & 04.2 (00.6) & 95.0 (05.2) & VLA 7\tablenotemark{a} & Class II\tablenotemark{a} \\ 
g & 18:30:09.65 -02:00:33.44 & 0.16 (0.03) & 0.25 (0.08) & 07.5 (01.9) & 04.5 (00.8) & 147.1 (13.0) & VLA 1\tablenotemark{a} & Extragal.\tablenotemark{a} \\ 
h & 18:30:01.56 -02:00:51.31 & 0.19 (0.04) & 0.31 (0.09) & 07.0 (01.5) & 05.1 (00.8) & 46.0 (20.1) & & \\ 
i & 18:30:15.03 -02:08:07.68 & 0.15 (0.03) & 0.26 (0.08) & 08.6 (02.4) & 04.2 (00.7) & 82.7 (08.5) & & \\ 
\enddata 
\tablerefs{$^{a}$ \citet{kern_2016}; $^b$ \citet{maury_2011}; $^c$ \citet{plunkett_2015_outflows}}
\end{deluxetable}
\end{rotatetable*}
\newpage

\section{\amm\ (3,3) fit results}

Here we show plots of the \amm\ (3,3) \vlsr\ and \sigv\ determined via fitting of a single Gaussian component as described in Section \ref{sec:line_fitting}. 

\begin{figure*}
    \centering
    \includegraphics[width=0.45\textwidth,trim={60 0 100 0},clip]{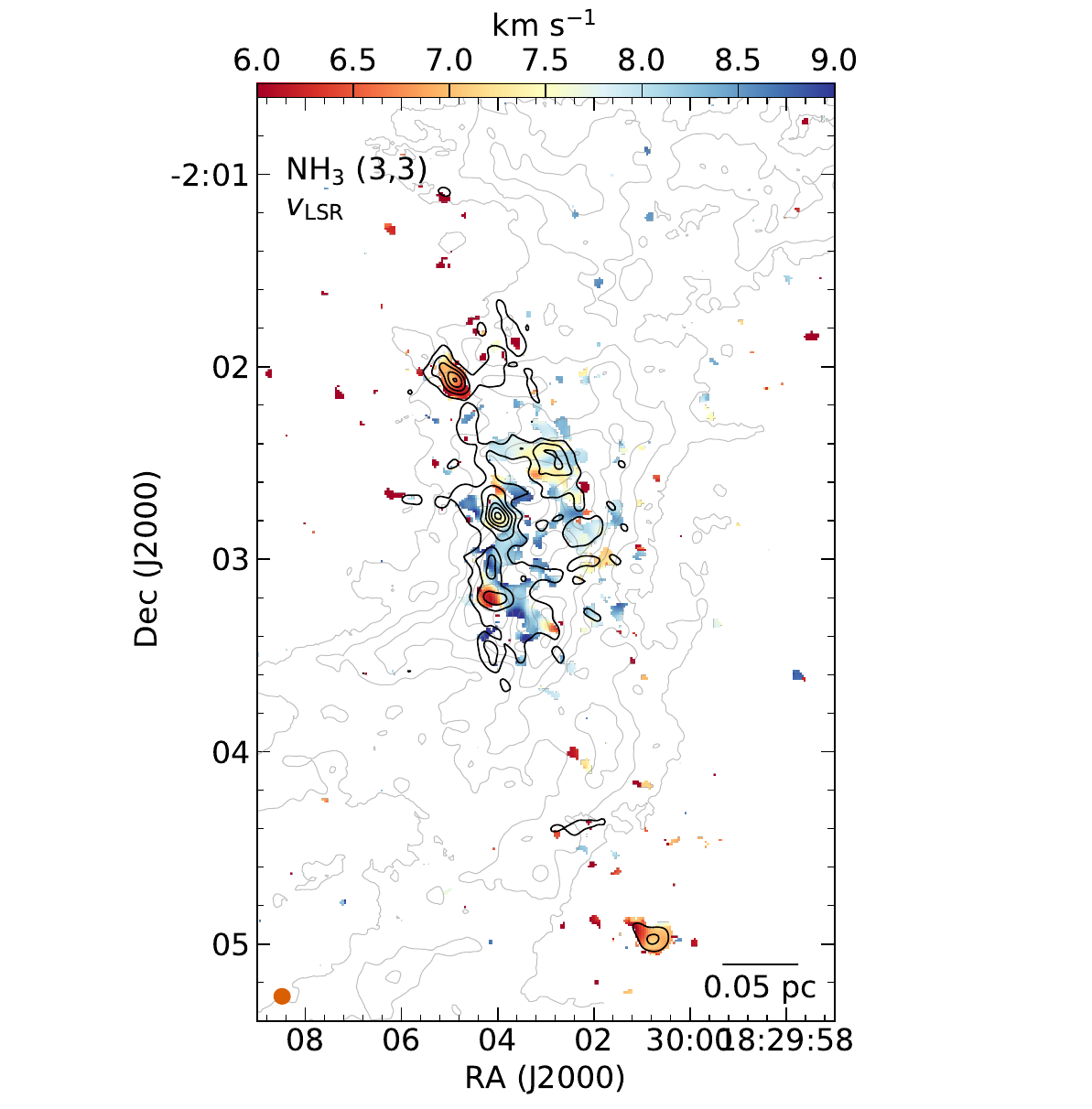}
    \includegraphics[width=0.45\textwidth,trim={60 0 100 0},clip]{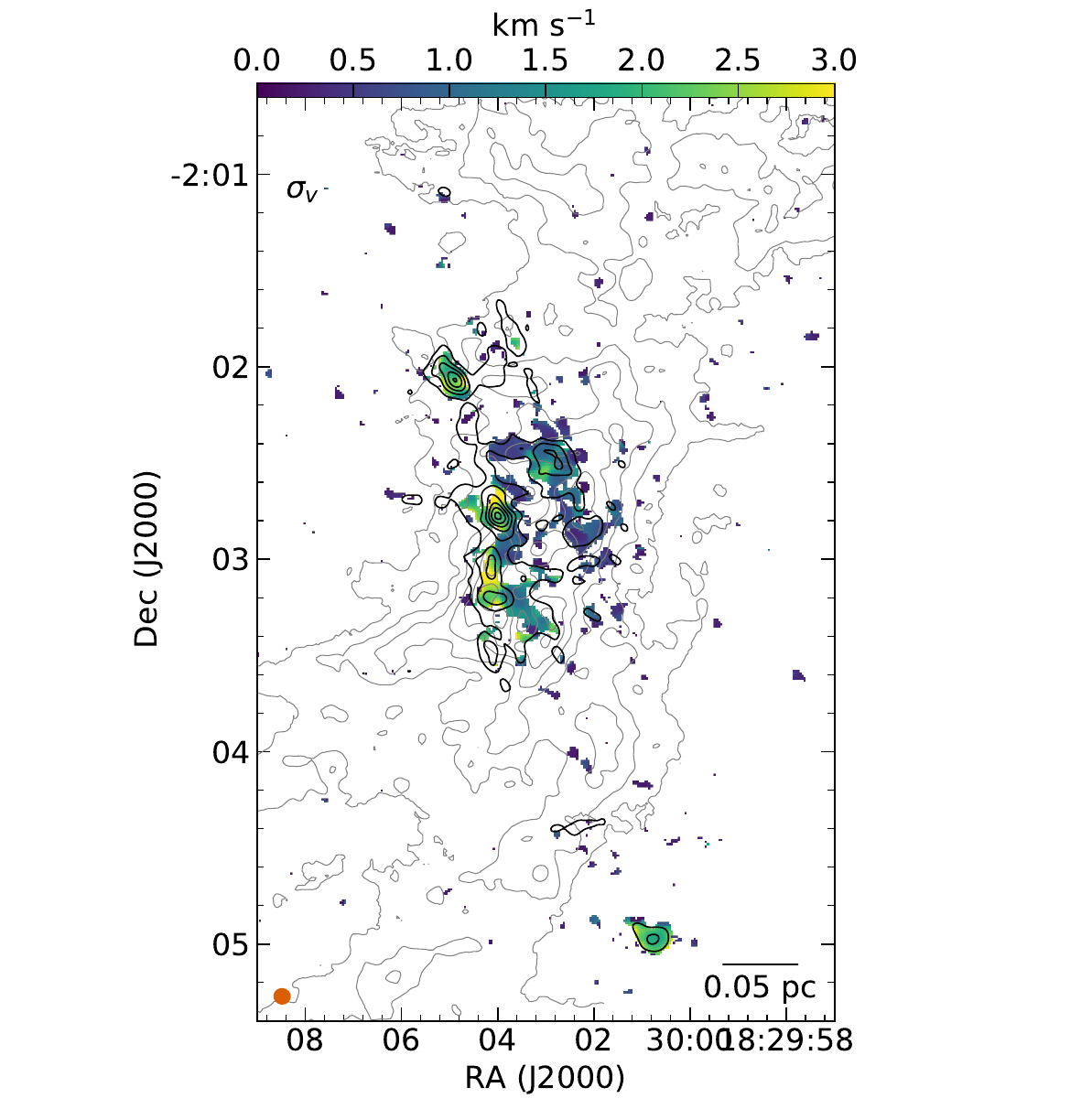}
    \caption{Single Gaussian fit results for \amm\ (3,3) \vlsr\ and \sigv. Black contours show \amm\ (3,3) integrated intensity as in Figure \ref{fig:cont_33}. Grey contours show \amm\ (1,1) integrated intensity as in Figure \ref{fig:protostars}. The beam is shown in orange at the bottom left. }
    \label{fig:nh3_33_fits}
\end{figure*}


\bibliography{biblio.bib}{}
\bibliographystyle{aasjournal}



\end{document}